\documentclass[11pt]{iopart}
\usepackage{comment}
\usepackage{enumerate}
\usepackage{amssymb}
\expandafter\let\csname equation*\endcsname\relax
\expandafter\let\csname endequation*\endcsname\relax
\usepackage{amsmath}
\usepackage{braket}
\usepackage{dsfont}
\usepackage{graphicx}
\usepackage[usenames,dvipsnames]{color}
\usepackage{tensor}
\usepackage{empheq}
\usepackage{capt-of}
\usepackage[normalem]{ulem} 
\usepackage{soul} %
\usepackage{MnSymbol}
\usepackage{cancel}

\usepackage[colorlinks,bookmarks=false,citecolor=NavyBlue,linkcolor=OliveGreen,urlcolor=blue]{hyperref}
\newcommand{\be}{\begin{equation}}
\newcommand{\ee}{\end{equation}}
\newcommand{\ba}{\begin{aligned}}
\newcommand{\ea}{\end{aligned}}
\newcommand{\bw}{\begin{widetext}}
\newcommand{\ew}{\end{widetext}}

\newcommand{\bea}{\begin{eqnarray}}
\newcommand{\eea}{\end{eqnarray}}

\newcommand{\II}{\mathrm{i}}
\newcommand{\D}{\mathrm{d}}

\def\doi{http://dx.doi.org/}

\begin{document}
\title{Entanglement gap in 1D long-range quantum spherical models 
}
\author{Sascha Wald$^{1,2}$, Raul Arias$^3$, and Vincenzo Alba$^4$
\footnote{Author to whom any correspondence should be addressed.}}
\address{$^1$Statistical Physics Group, Centre for Fluid and Complex Systems,
Coventry University, Coventry, England}
\address{$^2$\ $\mathbb{L}^4$ Collaboration \& Doctoral College for the
Statistical Physics of Complex Systems,
Leipzig-Lorraine-Lviv-Coventry, Europe}
\address{$^3$Instituto de F\'isica La Plata - CONICET and Departamento de
F\'isica, Universidad Nacional de La Plata C.C. 67, 1900, La Plata,
Argentina}
\address{
$^4$Dipartimento di Fisica dell' Universit\`a di Pisa and INFN, Sezione di Pisa, I-56127 Pisa, Italy}%
\ead{vincenzo.alba@unipi.it}

\begin{abstract}
We investigate the finite-size scaling of the entanglement gap in
the one-dimensional long-range quantum spherical model (QSM).
We focus on the \emph{weak} long-range QSM, for which
the thermodynamic limit is well-defined. 
This model exhibits a continuous phase transition, separating a paramagnetic from a ferromagnet
phase. 
The universality class of the transition
depends on the long-range exponent $\alpha$. We show that in the thermodynamic limit
the entanglement gap is finite in the paramagnetic phase, and it vanishes 
in the ferromagnetic phase. In the ferromagnetic phase the entanglement gap is 
understood in terms of standard magnetic correlation functions. 
The half-system entanglement gap decays as $\delta\xi\simeq C_\alpha L^{-(1/2-\alpha/4)}$,
where the constant $C_\alpha$ depends on the low-energy properties of the model 
and $L$ is the system size.
This reflects that the lower part of the dispersion is affected by the long range physics. 
Finally, multiplicative logarithmic corrections are absent in the scaling of the 
entanglement gap, in contrast with the higher-dimensional case. 
\end{abstract}

\maketitle

\section{Introduction}
\label{sec:intro}

\begin{figure}
\begin{center}
\includegraphics[width=0.75\linewidth]{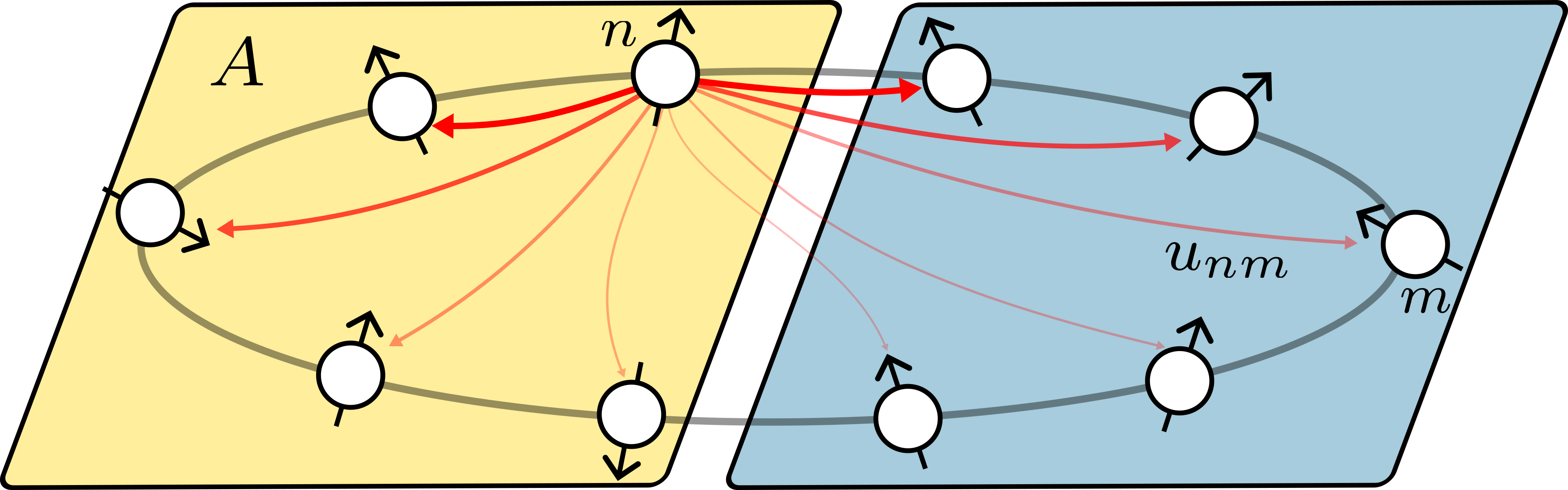}
\caption{ Schematic view of a one-dimensional spin system with long-range interactions
and periodic boundary conditions.
 Here $u_{nm}$ is the interaction potential between site $n$ and $m$. The system is
 translational invariant, i.e., $u_{nm}$ depends only on the distance $|n-m|$ 
 along the ring. The magnitude of $u_{nm}$ is depicted by the
 faintness of the arrow. The chain
 has $L$ sites and  periodic boundary conditions. 
 We are interested in the entanglement between a subsystem
 $A$ containing $L/2$ sites and the rest.
}
\label{fig:cartoon}
\end{center}
\end{figure}

In recent years, the investigation of entanglement patterns provided 
new insights into the structure of correlations in quantum many-body 
systems~\cite{amico2008entanglement,eisert2010colloquium,calabrese2009entanglement,laflorencie-2016}.
Here we focus on the 
so-called entanglement spectrum, which 
is one of the tools to investigate these quantum correlations and thus
has been the subject of intense activity in the last decade. 
The entanglement spectrum is derived from the entanglement Hamiltonian the definition
of which we now briefly recall.
Consider a one-dimensional quantum many-body systems that is 
prepared in the ground state $|\Psi\rangle$ of a Hamiltonian $H$. We 
divide the full system into two mutually exclusive parts
$A\cup B$ (see Fig.~\ref{fig:cartoon}) and consider the reduced density matrix 
$\rho_A:=\mathrm{Tr}_B|\Psi\rangle\langle \Psi|$ for the part $A$.
We define the entanglement Hamiltonian $\mathcal{H}_A$ by formally writing $\rho_A$ as
exponential, viz.,
\begin{equation}
	\rho_A=e^{-{\mathcal H}_A}.
\end{equation}
The eigenvalues $\xi_i$ of 
${\mathcal H}_A$ form the so-called entanglement spectrum (ES)
and are readily
given in terms of the eigenvalues $\lambda_i$ of $\rho_A$ as $\xi_i=-\ln(\lambda_i)$. 
The ES is a valuable tool to understand the performances of the Density Matrix 
Renormalization Group (DMRG) method~\cite{white-1992}, which triggered earlier studies~\cite{peschel-1999,peschel-2004}. 

Recently, the ES has been considered in fractional quantum Hall systems to study the
edge energy spectrum~\cite{thomale-2010,andreas-2010,haque-2007,thomale-2010a,hermanns-2011,chandran-2011,qi-2012,liu-2012,sterdyniak-2012,dubail-2012,dubail-2012a,Chan14}, 
%
%
%
in topological phases of matter~\cite{pollmann-2010,turner-2011,bauer-2014} or in systems 
that exhibit magnetic order~\cite{poilblanc-2010,cirac-2011,de-chiara-2012,alba-2011,metlitski-2011,alba-2012,Alba13,lepori-2013,james-2013,kolley2013entanglement,Chan14,rademaker-2015,kolley2015phase,frerot-2016,alba2021entanglement,contessi2022phase}.
Furthermore, the
ES also provides a versatile tool to understand the effects of impurities in quantum many-body 
systems~\cite{bayat-2014}. 
Interestingly, the framework of Conformal Field Theory (CFT)
allows one to obtain universal scaling properties of entanglement spectra 
analytically~\cite{calabrese2008entanglement,lauchli-2013,Alba-2017,cardy-talk,ruggiero-2016}. 

Despite the versatile use of the ES,
most of the literature focused on short-range models to date.
This has changed very recently with the
growing interest in models with long-range interactions~\cite{defenu2021long}, 
driven by the dramatic experimental progress~\cite{zhang2017observation}.
Concomitantly, there has been a rise in the interest in 
characterizing entanglement properties of long-range quantum many-body systems~\cite{koffel2012entanglement,Raja13,vodola2016long,frerot2017entanglement,gong2017entanglement,mozaffar2017entanglement,Moza22, maghrebi2017continuous,pappalardi2018scrambling,mozaffar2019entanglement,bentsen2022entanglement,ares2022symmetry}. 
Here we will consider one such model with long-range
interactions that allows to quantify a variety of entanglement properties 
analytically. Particularly, we focus on the entanglement gap $\delta\xi$, which is the
lowest laying gap of the entanglement Hamiltonian, i.e.,
\begin{equation}
\label{eq:gap-def}
\delta\xi=\xi_1-\xi_0,
\end{equation}
with $\xi_0$ and $\xi_1$ being the two lowest ES levels.
The entanglement gap received significant 
attention~\cite{truong-1987,peschel-1999,chung-2000,peschel-2004,alba-2011,andreas-2010,de-chiara-2012,lepori-2013,di-giulio-2020}. 
For instance, in CFT  systems $\delta\xi$ decays 
as $\delta\xi\propto 1/\ln(\ell)$ with $\ell$ the subsystem 
length~\cite{calabrese2008entanglement}. 
Similar results were also obtained by using the corner transfer matrix technique~\cite{truong-1987}.
In magnetically ordered phases of matter in $D>1$, which are associated with the 
breaking of a continuous symmetry,
the lower part of the ES bears a striking resemblance~\cite{metlitski-2011} 
to the Anderson tower-of-states~\cite{lhullier,beekman-2020,Wietek-2017}. 
Specifically, this implies that the entanglement gap exhibits a power-law decay 
as a function of the volume of the subsystem, with possible multiplicative logarithmic 
corrections. 
This prediction has been confirmed analytically in systems of quantum rotors~\cite{metlitski-2011}
and there is numerical evidence suggesting that this correspondence between ES and tower-of-states
structures is also present in the superfluid phase of the two-dimensional Bose-Hubbard
model~\cite{Alba13} (see also~\cite{frerot-2016}), and in two-dimensional 
Heisenberg antiferromagnets~\cite{kolley2013entanglement,kolley2015phase}. 

Interestingly, it was argued that in general the closure of
the entanglement gap is not associated with criticality~\cite{Chan14,lundgren-2014}.
Still, e.g. in the so-called spherical model~\cite{Ober72,Henk84,Vojta96,Wald15} in $2D$,
a closing of the gap is observed at criticality~\cite{Wald20-1}.
Here, the entanglement gap was even derived
analytically in Ref.~\cite{alba2021entanglement}
(see also~\cite{Wald20-1}).

Here we investigate the scaling of the entanglement gap in the ordered phase of 
one-dimensional long-range quantum many-body systems. 
We focus on the quantum spherical
model (QSM)~\cite{Ober72,Henk84,Vojta96,Wald15} with long-range couplings. 
The classical spherical model~\cite{Berl52} played a fundamental role in 
addressing the validity of Renormalization Group techniques~\cite{pelissetto2002critical} 
to describe critical phenomena. Its quantum version~\cite{Ober72,Henk84,Vojta96} provides a convenient 
framework to address the interplay of quantum and classical fluctuations at 
criticality. Quite generically, critical behavior in quantum and classical spherical models 
is in the universality class of the $O(N)$ 
vector model~\cite{zinn-justin-1998} with  $N\to\infty$~\cite{Stan68,Henk84,Vojta96}.
The $O(N)$ model and the spherical model are also valuable to investigate entanglement 
properties~\cite{Metlitski-2009,whit-2016,Lu19,Lu20,Wald20,Wald20-1,alba2021entanglement}. 
Here we consider the one-dimensional QSM with long range couplings. A pictorial view of the 
system is reported in Fig.~\ref{fig:cartoon}. In the presence of long-range couplings 
the model exhibits a second-order phase transition between a ferromagnetic phase and a standard 
paramagnetic one. The critical behavior depends on the 
the exact shape
of the long range interactions~\cite{Vojta96}. 

We consider a finite size system of length $L$ focusing on the bipartition into two 
parts $A$ and $B$ of equal length $L/2$ (see Fig.~\ref{fig:cartoon}). 
We show that the entanglement gap is finite in the paramagnetic phase 
and remains finite in the thermodynamic limit $L\to\infty$, whereas it vanishes in the 
ferromagnetic phase. In the ferromagnetic phase, 
the decay of the entanglement gap follows a
power-law as $\delta\xi\simeq C_\alpha L^{-1/2-\alpha/4}$. 
Here $C_\alpha$ is a constant that depends only on the low-energy properties of the 
model. Interestingly, in the ferromagnetic phase the entanglement gap is 
directly related to the magnetic correlation functions $\chi_A^x$ and $\chi_A^t$. 
Here $\chi_A^x$ is the susceptibility associated with the spherical coordinate degrees of freedom. 
On the other hand, $\chi_A^t$ is the susceptibility associate with the momentum-like 
conjugate variable. 
In the ordered phase $\chi_A^x\simeq L$ which reflects
that despite the presence of the long-range terms, the structure of correlations in 
the ground state is the standard one for a ferromagnet. The susceptibility $\chi_A^t$ 
contains information about the low-energy part of the dispersion, and hence on the 
long-range terms. Indeed, the dependence on 
the shape of the long-range interactions
in the ES originates
from $\chi_A^t$. Precisely, in the ferromagnetic phase we show that $\chi_A^t\simeq L^{-\alpha/2}$. 
Hence, $\chi_A^t$ vanishes in the thermodynamic limit. The prefactor, which we determine analytically, 
depends only on the singular behavior of the dispersion, and not on the high-energy part. 

The paper is organized as follows. In section~\ref{sec:model} we introduce the one-dimensional 
QSM. We discuss its behavior at criticality and in the ordered phase. In particular, we derive 
analytically the finite-size scaling of the spherical parameter, which to the best of our 
knowledge was not known. In section~\ref{sec:ent-def} we briefly review how to extract the 
entanglement spectrum and the entanglement gap. In section~\ref{sec:main} we outline the 
derivation of our main result. Section~\ref{sec:numerics} is devoted to numerical 
benchmarks. We discuss some future directions in section~\ref{sec:concl}. 

In~\ref{app:gc} we derive the critical coupling marking the second-order phase 
transition as a function of the long-range exponent $\alpha$. In~\ref{app:mu} we 
derive the finite-size scaling behavior of the spherical parameter both  at criticality 
and in the ordered phase. In~\ref{app:flatX} and~\ref{app:pcorr} we derive the finite-size 
scaling behavior of $\chi_A^x$ and $\chi_A^t$, respectively.

%
%

\section{Quantum Spherical Model (QSM) with long-range interactions}
\label{sec:model}

The spherical model~\cite{Berl52} was originally introduced as a simplification of the Ising model, 
and has established itself as a reference system to investigate collective properties of 
strongly-interacting systems. Indeed, the spherical model allows for analytical 
investigation of many-body systems beyond mean-field transitions.

In its quantum formulation, the QSM becomes equivalent  to a system of harmonic oscillators
subject to a single global constraint. The Hamiltonian of the one-dimensional QSM
with periodic boundary conditions is~\cite{Ober72,Henk84,Vojta96,Wald15}
\begin{equation}
\label{eq:qsm-ham}
 H = \sum_{n = 1}^L  \left[ \frac{g}{2} p_n^2 +
 \frac{1}{2}\sum_{m=1}^L u_{nm} x_n x_m \right].
\end{equation}
The operators $x_n$ and $p_n$ are the conjugated oscillator position and momentum
operators, satisfying the canonical commutation relation
$[x_n,p_m] = \II \hbar \delta_{nm}$. The oscillators
interact through the translation invariant potential
$u_{nm} = u(|n-m|)$. To decouple the oscillators, we introduce the Fourier transformed 
operators $q_k,\pi_x$ as
\begin{align}
 x_{n} = \frac{1}{\sqrt{L}}
	\sum_{k\in \mathcal{B}} e^{\II k n }q_{k}, \qquad
	p_{n} = \frac{1}{\sqrt{L}}
	\sum_{k\in \mathcal{B}} e^{-\II  k n } \pi_k\ ,
\end{align}
with the Brillouin zone $\mathcal{B}=\{0,2\pi/L,...,2\pi(L-1)/L\}$.
The Hamiltonian in Eq.~\eqref{eq:qsm-ham} then reads
\begin{equation}
	\label{eq:qsm-ham-1}
	H
=\sum_{k\in \mathcal{B}}  \left[ \frac{g}{2}  \pi_k \pi_{-k}
+   \frac{1}{2}u(k)q_{k} q_{-k} \right],
\end{equation}
with $u(k)$ the Fourier transformed interaction potential. 
For nearest-neighbor interactions, $u(k)$ is a discretized Laplacian,
i.e., $u(k) = 2\mu+2(1-\cos k)$. It has been argued that
long-range interactions may be introduced by replacing the
Laplacian by its fractional counterpart~\cite{Zoi07,Raja13} as 
\begin{align}
\label{eq:lap-frac}
 u(k) = 2\mu+ \left(2(1-\cos k)\right)^{\frac{\alpha}{2}}.
\end{align}
Indeed, in real space, Eq.~\eqref{eq:lap-frac} corresponds  to the interaction
potential
\begin{equation}
 u(|n-m|)
 \stackrel{|n-m|\to\infty}{\simeq}
 -\frac{\Gamma\left( 1+\alpha\right)}{\pi} \sin\left(\frac{\alpha}{2}\pi\right)
 \left(\frac{1}{|n-m|}\right)^{\alpha+1}, 
\end{equation}
which is clearly long-range. The strength of the interaction is parametrized by 
the long-range exponent $\alpha$.
Here we consider $0<\alpha<2$, such that the 
interaction potential satisfies the condition $1+\alpha > d =1$.
In this regime, which is sometimes referred to as
{\it weak long-range} regime, the thermodynamic limit is
well-defined as the interactions decay sufficiently
fast with distance~\cite{defenu2021long}.
The parameter $\mu$ is a Lagrange parameter chosen {\it self-consistently} to 
ensure the spherical constraint as~\cite{Berl52,Vojta96,Bien12,Wald15}
\begin{equation}
\label{eq:constraint}
\sum_{n=1}^L \left<x_n^2\right> = L.
\end{equation}
This constraint distinguishes the QSM from a 
simple collection of harmonic oscillators, and is responsible for supporting a
quantum phase transition at zero temperature. To pinpoint this transition,
we diagonalize the Hamiltonian in Eq.~\eqref{eq:qsm-ham-1} by introducing bosonic ladder operators
$b_k,b^\dagger_k$ as 
\begin{equation}
\label{eq:qk}
 q_k = \alpha_k \frac{b_k+b_{-k}^\dagger}{\sqrt{2}},
 \qquad
 \pi_k = \frac{\II}{\alpha_k} \frac{b_k^\dagger  - b_{-k}}{\sqrt{2}}, 
\end{equation}
with $\alpha_k^4 = g/u(k)$ \cite{Wald15}. Hence, the Hamiltonian $H$ becomes 
diagonal and Eq.~\eqref{eq:qsm-ham-1} can be written as 
\begin{equation}
	\label{eq:ham-diag}
	H = \sum_{k\in\mathcal{B}} E_k\left(b_k^\dagger b_k +\frac{1}{2}\right),\qquad E_k:=\sqrt{gu(k)}.
\end{equation}
To determine the critical behavior of the QSM at zero temperature
and to study entanglement properties (see section~\ref{sec:ent-def}), it is necessary to obtain the position and momentum correlation functions 
$\mathbb{X}_{nm}$ and $\mathbb{P}_{nm}$ respectively. A straightforward calculation gives~\cite{Wald15}
\begin{subequations}
\label{eqs:corr}
\begin{align}
\label{eq:xx}
 \mathbb{X}_{nm}&:= \langle x_{n} x_{m} \rangle= \frac{g}{2L}
 \sum_{k} e^{\II (n-m)k}  \frac{1}{E_k}, \\
 \label{eq:pp}
\mathbb{P}_{nm}&:=\langle p_{n} p_{m} \rangle=\frac{1}{g} \frac{1}{2L}\sum_{k}
e^{-\II (n-m)k}  E_k,
\end{align}
\end{subequations}
where $\langle\cdot\rangle$ denotes the ground-state expectation value. 
In the thermodynamic limit $L\to\infty$ the diagonal components of
the correlator $\mathbb{X}_{nm}$ allow to rewrite the spherical 
constraint (cf. Eq.~\eqref{eq:constraint}) as 
\begin{equation}
\label{eq:sc}
\frac{2}{g} = \frac{1}{L} \sum_k \frac{1}{E_k}
 \stackrel{L\to\infty}{\to}
 \int_0^{2\pi} \frac{\D k}{2\pi} \frac{1}{E_k}.
\end{equation}
In the thermodynamic limit Eq.~\eqref{eq:sc} has a finite 
solution $\mu>0$ as long as the tuning parameter $g$ satisfies $g>g_c$.
Conversely, for $g\le g_c$ one finds that $\mu$ is 
identically zero. The nonanalytic behavior of $\mu$ as a function 
of $g$ determines the  critical properties of the model. The quantum 
critical point at $g_c$ marks the transition between a paramagnetic phase 
at $g>g_c$ and a ferromagnetically ordered one at $g<g_c$. 
The critical coupling $g_c$ is obtained by imposing the condition
$\mu=0$~\cite{Vojta96,Wald15}. Direct integration of the constraint
then yields (see~\ref{app:gc})
\begin{equation}
g_c=
2^{\alpha+2}\pi
\left(\frac{\Gamma\left(1-\alpha/4\right)}{
\Gamma\left(1/2-\alpha/4\right)}\right)^{2}.
\label{eq:gc}
\end{equation}
\begin{figure}[t!]
\begin{center}
\includegraphics[width = 0.6\columnwidth]{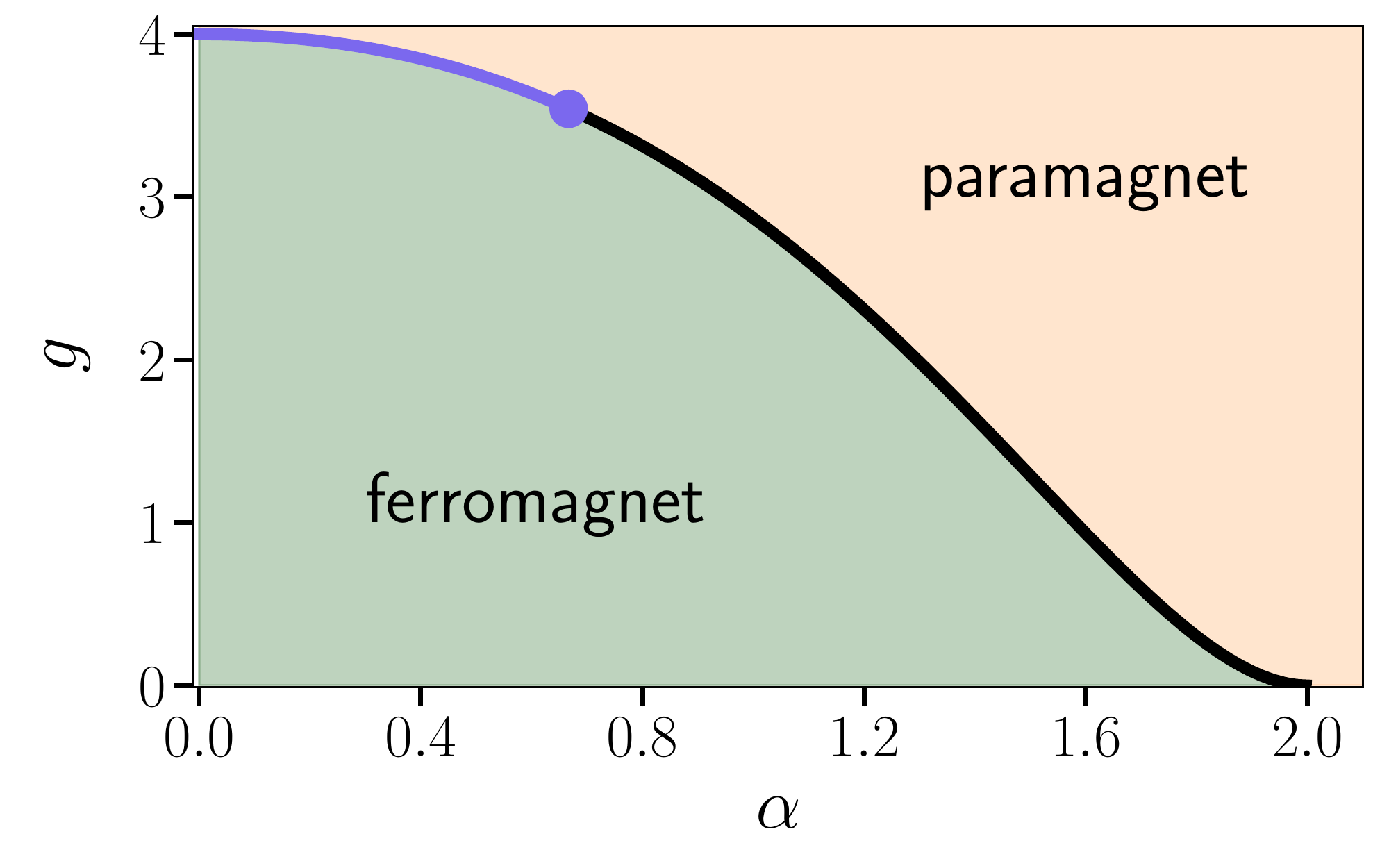}
\caption{Zero-temperature phase diagram of the quantum spherical model (QSM) with
 long-range interactions. The plot shows the critical coupling $g_c$
 as a function of the decay exponent  $\alpha$  of the long-range interactions (continuous line).
 Here we restrict ourselves to $0\le \alpha< 2$, i.e., to the regime of
 \emph{weak} long-range  interactions, for which the thermodynamic limit is well defined. 
 At $g=g_c$ the QSM exhibits a second-order quantum phase transition, which divides
 a paramagnetic phase from a ferromagnetically ordered one. For $\alpha\geq 2$
 interactions are effectively short-ranged, and the QSM is not critical.
 For $\alpha\leq 2/3$ (dot in the figure) the transition is of mean-field type.
}
\end{center}
\label{fig:pd}
\end{figure}
The resulting zero-temperature phase diagram 
is shown in Fig.~\ref{fig:pd} for $0\le\alpha\le2$.
Notice that 
for $\alpha>2$ the model becomes effectively short range, and the 
critical behavior disappears, as expected for a one-dimensional model. 
One can also show that for $0\le \alpha\le 2/3$ the phase transition is of 
mean-field type, see Ref.~\cite{Vojta96} or \ref{app:mu} for further details. 
Thus, at least for $2/3<\alpha<2$, the QSM supports non-mean-field criticality despite being a 
Gaussian system. This is due to the nontrivial spherical constraint, see Eq.~\eqref{eq:constraint}. 

Let us now discuss the finite-size scaling of the spherical parameter $\mu$. For finite 
$L$ Eq.~\eqref{eq:constraint} gives a nonzero value of $\mu$ for any $g$. Upon increasing $L$, the 
spherical parameter 
$\mu$ retains a finite value for $g>g_c$, whereas it vanishes for $g\le g_c$. The precise 
behaviors of $\mu$ at the critical point $g_c$ and in the ordered phase  are different. 
Specifically, in~\ref{app:mu} we show that the finite-size scaling of $\mu$ is given by
\begin{equation}
	\label{eq:mu-fss}
	\mu = \begin{dcases}
	\quad\frac{\gamma_\alpha}{L^{\alpha}}, &g = g_c\\
	\quad\frac{1}{8}\left(\frac{1}{\sqrt{g}}-\frac{1}{\sqrt{g_c}}\right)^{-2}\frac{1}{L^2}, &g<g_c. 
	\end{dcases}
\end{equation}
%
%
%
\begin{figure}[t]
\centering
\includegraphics[width=.6\textwidth]{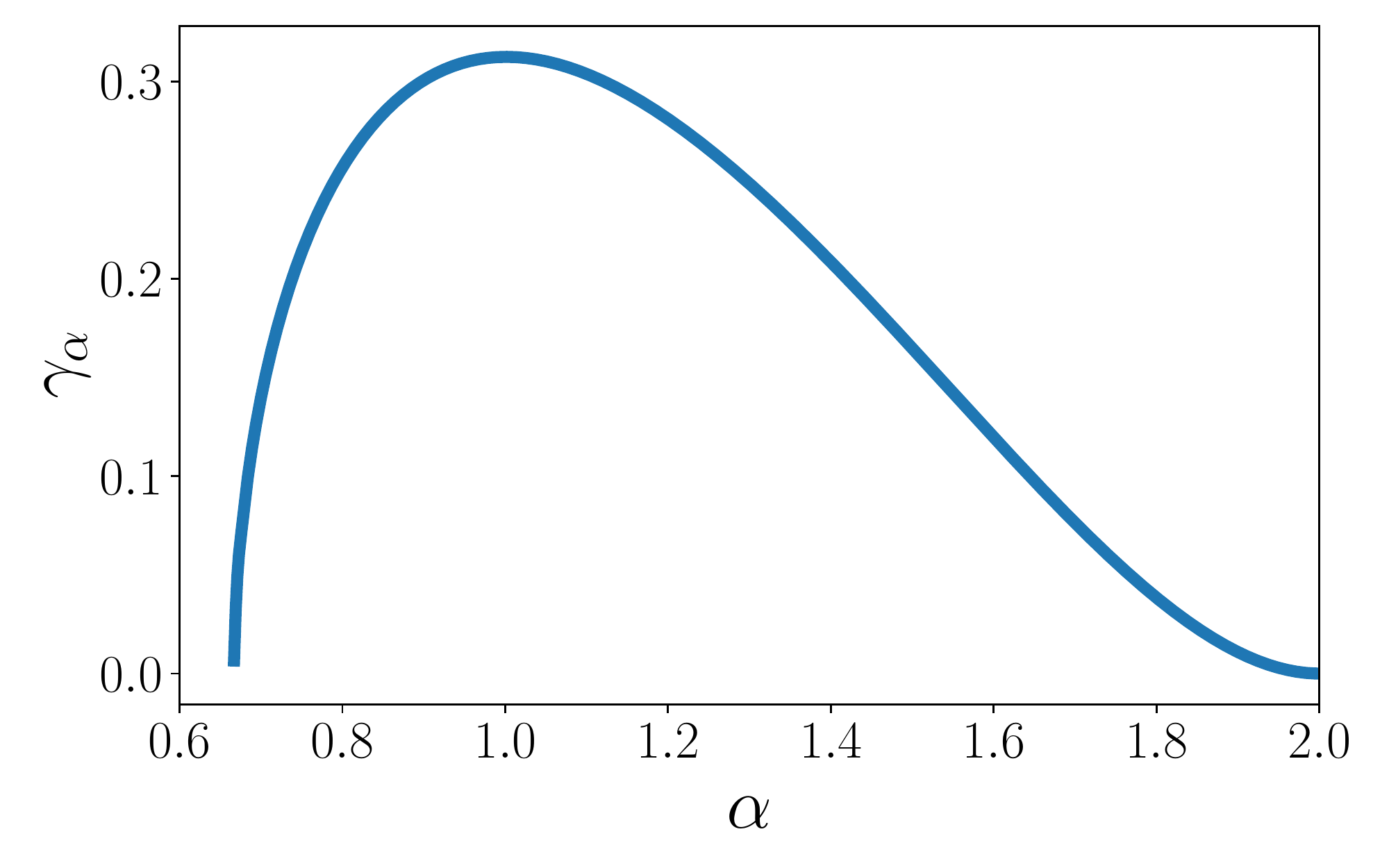}
\caption{ Prefactor $\gamma_\alpha$ of the finite-size scaling behavior of the 
 spherical parameter $\mu=\gamma_\alpha/L^\alpha$ at the critical point. Here 
 we plot $\gamma_\alpha$ versus the exponent $\alpha$ of the long-range interactions. 
 We only consider the region $2/3<\alpha< 2$. Notice the vanishing behavior for $\alpha\to2$ and $\alpha\to2/3$. 
 For $\alpha\to2$ the model becomes short range and there is no critical behavior. 
 For $\alpha\le 2/3$  the  transition becomes of the mean-field type. 
 The curve is obtained by numerically solving Eq.~\eqref{eq:mu-c-eq-main}. 
}
\label{fig:gamma-theo}
\end{figure}
%
In Eq.~\eqref{eq:mu-fss} we show only the leading behavior of $\mu$ in 
the limit $L\to\infty$. Notice that deep in the ferromagnetic phase, i.e., for 
$g\ll g_c$, Eq.~\eqref{eq:mu-fss} yields $\mu\simeq g/(8L^2)$. 
The scaling for $g<g_c$ is determined solely by the zero mode at $k=0$ in 
the dispersion $E_k$ (cf. Eq.~\eqref{eq:ham-diag}). 
Notice that from $\mu$ one can define the correlation length $\xi_\mathrm{corr}$ 
of the QSM~\cite{Vojta96} as $\xi_\mathrm{corr}=\mu^{-1/\alpha}$. 
The constant $\gamma_\alpha$ in Eq.~\eqref{eq:mu-fss} is universal, and is obtained by solving the 
equation (see~\ref{app:mu}) 
\begin{equation}
	\label{eq:mu-c-eq-main}
	\pi^{-\frac{3}{2}}\Gamma\left(\frac{1}{2}-\frac{1}{\alpha}\right)\Gamma\left(1+\frac{1}{\alpha}
	\right)(2\gamma_\alpha)^{\frac{1}{\alpha}-\frac{1}{2}}+(2\gamma_\alpha)^{-\frac{1}{2}}
	+4\gamma_\alpha^{\frac{1}{\alpha}-\frac{1}{2}}r'+4\sum_{k=0}^\infty\gamma_\alpha^k r_k=0, 
\end{equation}
with $r_k$ given by 
\begin{equation}
\label{eq:t3-main}
	r_k:=
	\frac{(-1)^k }{k!}
	\frac{2^{k-1}}{\pi^{\frac{3}{2}}}
	\Gamma \left(k+\frac{1}{2}\right)
  \sin \left(\frac{\pi}{4} 
   \alpha  (2 k+1)\right) \Gamma \left(1-k \alpha -\frac{\alpha
   }{2}\right) \zeta \left(1-\frac{\alpha}{2} (2 k+1)  \right), 
\end{equation}
and $r'$ defined as 
\begin{equation}
	\label{eq:r-def}
r'=-2^{\frac{1}{\alpha}-\frac{5}{2}}\pi^{-\frac{3}{2}}
\Gamma\left(\frac{1}{2}-\frac{1}{\alpha}\right)\Gamma\left(1+\frac{1}{\alpha}\right).
\end{equation}
In Eqs.~\eqref{eq:t3-main} and~\eqref{eq:r-def} $\Gamma(x)$ is the Euler gamma function, and 
$\zeta(x)$ is the Riemann zeta function. 
Importantly, Eq.~\eqref{eq:mu-c-eq-main} holds 
only in the region $2/3<\alpha<2$, in which the critical behavior is not of mean-field 
type. For $\alpha\to2/3$ and $\alpha\to2$, $\gamma_\alpha$ vanishes, and it exhibits 
a maximum at $\alpha\approx 1$. 
One should also notice that Eq.~\eqref{eq:mu-c-eq-main} depends on an infinite 
number of constants $r_p$. Still, it is straightforward to check that $r_p$ decays 
exponentially with increasing $p$, which implies that one can effectively truncate 
the sum in~\eqref{eq:mu-c-eq-main}. 
We show $\gamma_\alpha$ as a function of $\alpha$ in Fig.~\ref{fig:gamma-theo}. 
The continuous line is obtained by numerically solving~\eqref{eq:mu-c-eq-main}. 
Again, our results hold for $\alpha>2/3$, although they could be straightforwardly 
generalized to the mean field region $\alpha\le 2/3$. 
Moreover, we numerically observed that in the mean-field region (see Fig.~\ref{fig:pd}) 
 $\mu$ still decays as a power law in the large $L$ limit, although 
we did not extract the precise finite-size scaling behavior. 

Importantly, both at criticality and 
in the ferromagnetic phase the scaling of $\mu$ at leading order for large  $L$ 
depends only on the low-energy properties of the model. 
Finally, it is interesting to observe that 
for $\alpha=1$, the critical exponents of the QSM become the same 
as those of the two-dimensional short-range QSM. Still, 
the constant $\gamma_1$ is not expected to be the same in the 
two models, because $\gamma_\alpha$ depends 
on the dimensionality and boundary conditions. 
%
%
\begin{figure}[t]
\centering
\includegraphics[width = 0.495\columnwidth]{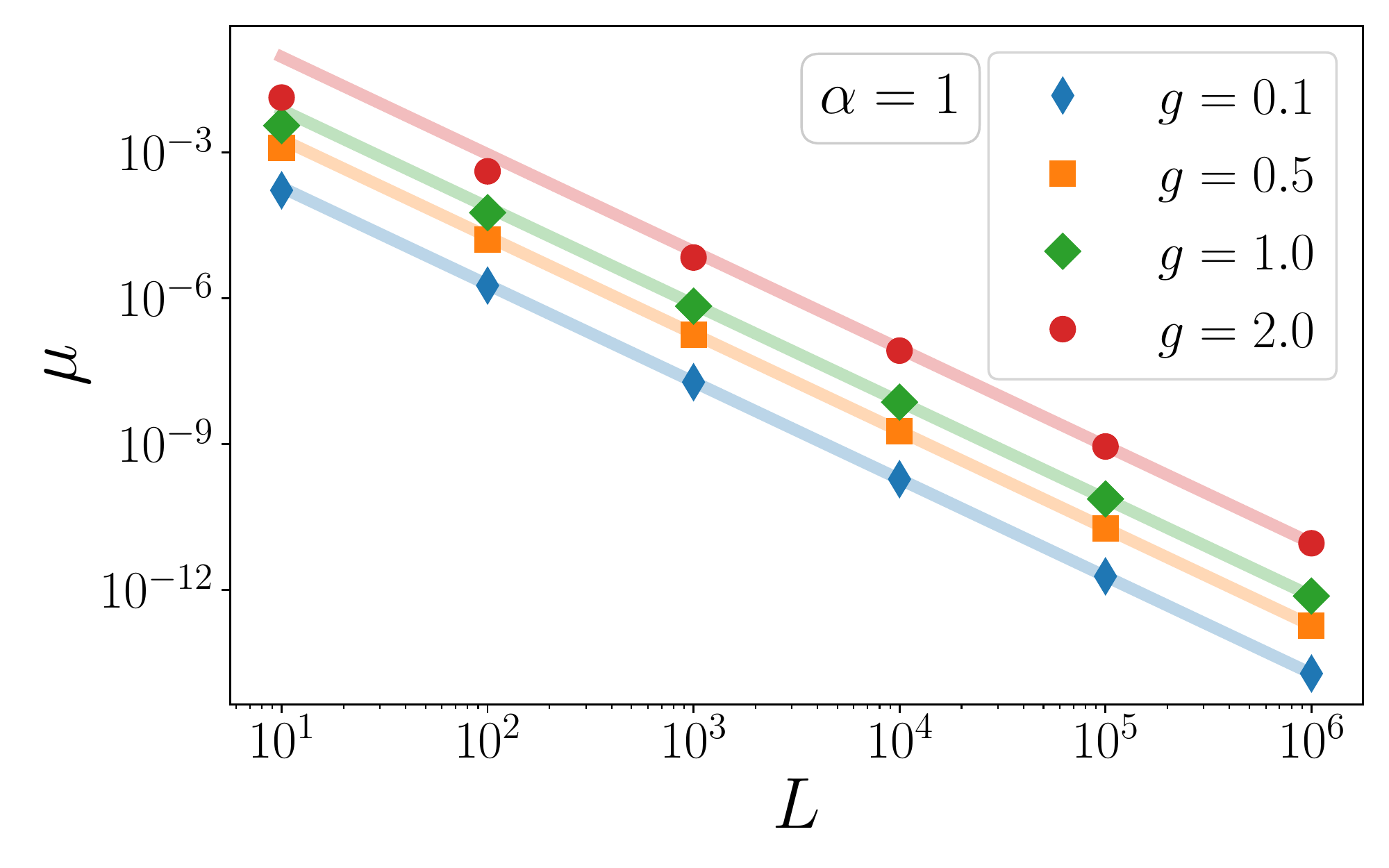}
\includegraphics[width = 0.495\columnwidth]{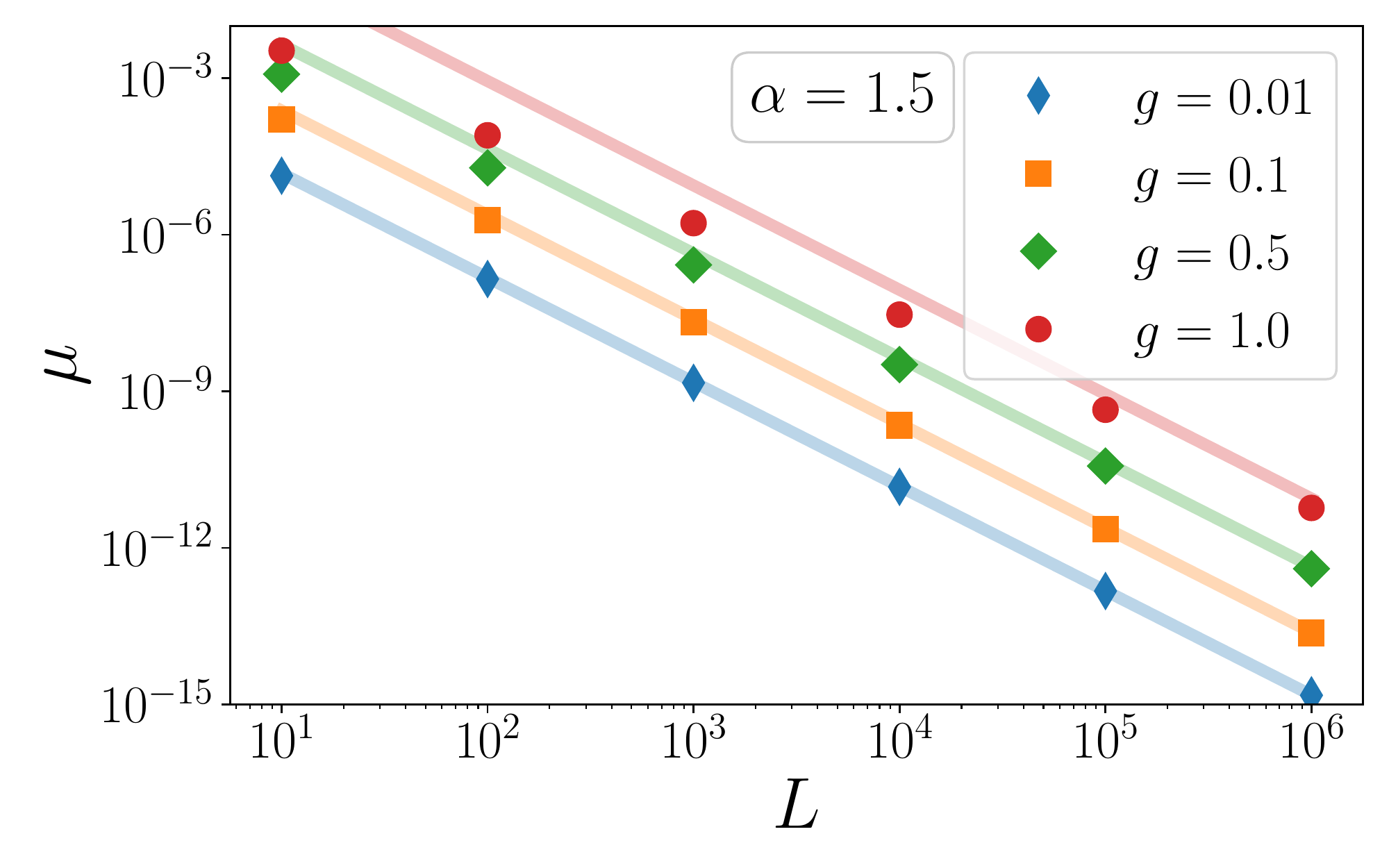}
\caption{ Finite-size scaling of the spherical parameter $\mu$ in the QSM 
 with long-range interactions. We show $\mu$ plotted versus $L$ for $\alpha=1$ and 
 $\alpha=1.5$ (in the left and right panel, respectively). The different symbols correspond to 
 different value of the coupling $g$. All the results are for the ferromagnetic phase at $g<g_c$. 
 The continuous lines are the analytical results for $L\to\infty$ (cf.~\eqref{eq:mu-fss}). 
}
\label{fig:mu}
\end{figure}
%

In Fig.~\ref{fig:mu} we numerically  verify the finite-size
scaling of the spherical parameter (cf. Eq.~\eqref{eq:mu-fss}) in the ferromagnetic 
phase. Specifically, in the figure we show numerical results for $\mu$ as a function 
of $L$, obtained 
by solving Eq.~\eqref{eq:constraint}. 
The left and right panels show results for $\alpha=1$ and $\alpha=3/2$, respectively. 
In both cases $\mu$ decays as a power-law in the limit $L\to\infty$ (notice the logarithmic 
scale on both axes). 
In each panel, the different symbols correspond to different values of the coupling $g$. 
The continuous lines are the analytic results in Eq.~\eqref{eq:mu-fss}, and are 
in agreement with the numerical data in the limit $L\to\infty$. The agreement is 
perfect deep in the ferromagnetic phase. Finite-size corrections increase upon 
approaching the critical point, which signals the different scaling as $L^{-\alpha}$ 
at criticality. As it is clear from Fig.~\ref{fig:mu}, 
upon approaching criticality, larger system sizes are needed to observe the 
asymptotic scaling predicted in Eq.~\eqref{eq:mu-fss}. 

\begin{figure}[t]
\centering
\includegraphics[width=.6\textwidth]{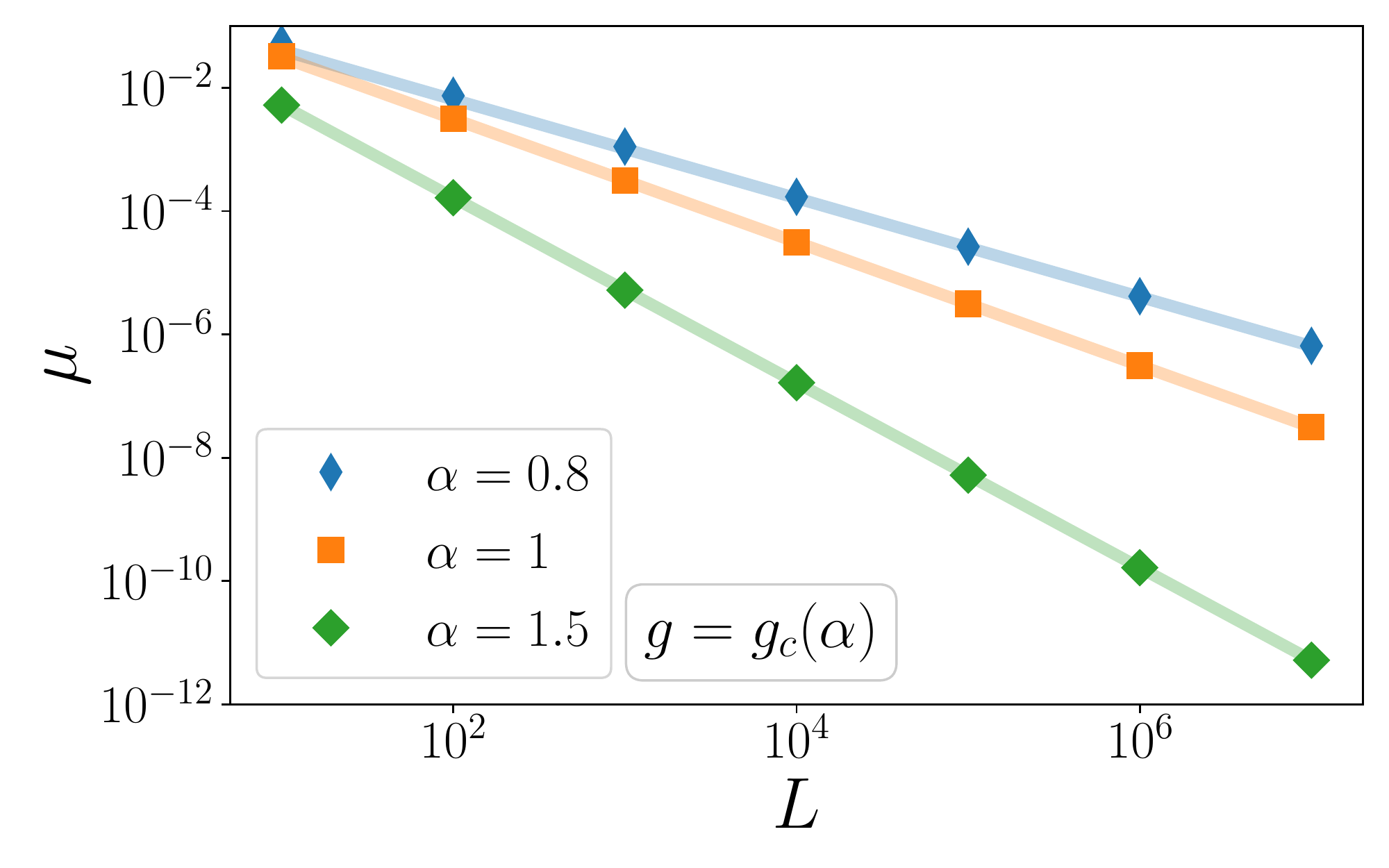}
\caption{ Finite-size scaling of the spherical parameter $\mu$ in the 
 critical long-range QSM: $\mu$ is plotted versus the system size $L$. 
 Different symbols are for different values of the exponent $\alpha$ of the long-range 
 interactions. Here we only consider the case $2/3<\alpha<2$, in which the 
 critical behavior is not of mean-field type. The continuous lines denote 
 the analytic result $\gamma_\alpha/L^\alpha$, with $\gamma_\alpha$ obtained 
 by solving Eq.~\eqref{eq:mu-c-eq-main}. 
}
\label{fig:gamma-num}
\end{figure}

Let us now discuss the finite-size scaling of $\mu$ at the phase transition (continuous line 
in Fig.~\ref{fig:pd}). Again, we focus on the region $2/3<\alpha<2$, i.e., where the transition 
is not of  mean-field type. Fig.~\ref{fig:gamma-num} shows numerical results for $\mu$ 
plotted as a function of $L$. Different symbols correspond to different values of 
the long-range exponent $\alpha$. The continuous lines are the analytical
predictions from Eq.~\eqref{eq:mu-fss}, 
with $\gamma_\alpha$ obtained by solving~\eqref{eq:mu-c-eq-main} 
(see Fig.~\ref{fig:gamma-theo}). The agreement between the numerical data and the analytical 
results is perfect. We anticipate that the finite-size scaling of $\mu$ presented here will be useful in 
section~\ref{sec:main} to determine the finite-size scaling of the entanglement gap. 
%
%
%

%
%
\section{Entanglement properties of the QSM}
\label{sec:ent-def}
Here we summarize the calculation of entanglement-related quantities in the QSM. 
As discussed in section~\ref{sec:model}, the QSM is mappable to a system of 
free bosons with the global spherical constraint, see Eq.~\eqref{eq:constraint}.
This ensures that entanglement related properties can be computed from 
the bosonic correlation functions~\cite{viktor}. 
Specifically, the reduced density matrix $\rho_A$ of a generic 
subregion $A$ (see Fig.~\ref{fig:cartoon}) for a system of free bosons 
can be written as~\cite{viktor}
\begin{equation}
\rho_A=Z^{-1}e^{-{\cal{H}}_A},\qquad
{\cal{H}}_A=\sum_k\epsilon_k b_k^\dagger  b_k.\label{rdm}
\end{equation}
with  ${\cal{H}}_A$ the entanglement Hamiltonian, $\epsilon_k$ 
the single-particle entanglement spectrum (ES), and $b_k$, $b_k^\dagger$ 
the bosonic ladder operators introduced in Eq.~\eqref{eq:qk}. The constant $Z$ ensures the normalization
of $\rho_A$ such that $\mathrm{Tr}(\rho_A) = 1$. The single-particle ES levels $\epsilon_k$ are  
readily related to the eigenvalues of the correlation matrix because the QSM is Gaussian. Again, 
entanglement properties of Gaussian systems are encoded in the two-point correlation matrices. 
For free bosons 
one has to compute the matrices~\eqref{eq:xx} and~\eqref{eq:pp}, where the chemical potential $\mu$ is 
self-consistently determined from Eq.~\eqref{eq:constraint}. To proceed, one has to 
compute the restricted correlation matrix $\mathbb{C}_A$, which is defined as 
\begin{equation}
	\label{eq:Corr-def}
	\mathbb{C}_A :=\mathbb{X}_A \cdot\mathbb{P}_A, \quad \mathbb{X}_A(\mathbb{P}_A)=
	\mathbb{X}_{ij}(\mathbb{P}_{ij})\,\,\mathrm{with}\,\, i,j\in A. 
\end{equation}
The entanglement spectrum and the eigenvalues $\epsilon_k$ are related to the eigenvalues 
$e_k$ of $\mathbb{C}_A$ as~\cite{viktor}
\begin{equation}
	\sqrt{e_k}=\frac{1}{2}\coth\left(\frac{\epsilon_k}{2}\right). 
\label{eq:ES}
\end{equation}
The ES of the QSM is then obtained by populating the single-particle levels
$\epsilon_k$ (cf.~\eqref{eq:ES}). We find
%
\begin{equation}
	\label{eq:ES-level}
	\xi\left(\{\beta_k\}\right)=\ln(Z)+\sum_{j}\beta_j\epsilon_j. 
\end{equation}
Here $\beta_k\in\mathbb{N}$ is the number of bosons in the 
single-particle ES level $\epsilon_k$, and $Z$ is the same normalization 
factor as in~\eqref{rdm}, viz.,
\begin{equation}
	\label{eq:norm}
	Z=\prod_{j=1}^{|A|}\left(\sqrt{e_j}+1/2\right), 
\end{equation}
where $|A|$ is the size of $A$. 
The lowest ES level corresponds to $\beta_j=0$ for any $j$. 
Let us assume that the single-particle ES levels are ordered as $\epsilon_1<\epsilon_2<\cdots<\epsilon_{|A|}$. The first excited ES level is obtained by populating the smallest single particle level
$\epsilon_1$. Thus, the lowest entanglement gap $\delta\xi$ (Schmidt gap) is defined as
\begin{equation}
	\label{eq:ES-gap}
	\delta\xi=\xi_1-\xi_0=\epsilon_1, 
\end{equation}
and $\epsilon_1$ is related to the eigenvalue $e_1$ of $\mathbb{C}_A$ via Eq.~\eqref{eq:ES}. 

\section{Finite-size scaling of the entanglement gap in the ordered phase of the long-range QSM}
\label{sec:main}

Our main result is that in the ordered phase of the long-range QSM (see Fig.~\ref{fig:pd}) 
the  eigenvalue $e_1$ of the restricted correlation matrix $\mathbb{C}_A$ 
(cf. Eq.~\eqref{eq:Corr-def}) in the large $L$ limit scales as 
\begin{equation}
	\label{eq:e1chi}
	e_1=\chi_A^x\chi_A^t, 
\end{equation}
where $\chi_A^{x,t}$ are the coordinate  and momentum ``susceptibilities'' defined as 
\begin{equation}
	\label{eq:su-def}
	\chi_A^x:=\langle1|\mathbb{X}|1\rangle_A,\quad \chi_A^t:=\langle1|\mathbb{P}|1\rangle_A. 
\end{equation}
Here $\mathbb{X}$ and $\mathbb{P}$ are defined in Eqs.~\eqref{eq:xx} and~\eqref{eq:pp}, respectively.  
Moreover, we introduced the normalized flat vector 
$|1\rangle:=(1,1,\cdots,1)/\sqrt{L_A}$ restricted to  subsystem $A$. 
The expectation values in Eq.~\eqref{eq:su-def} are defined as 
\begin{equation}
\langle1|\mathbb{X}(\mathbb{P})|1\rangle_A:
=\frac{1}{L_A}\sum_{n,m=1}^{L_A} \mathbb{X}_{nm}(\mathbb{P}_{nm}). 
\end{equation}
To proceed, it is crucial to observe that for $g<g_c$ the system develops ferromagnetic order, 
for any value of $\alpha<2$. This is reflected in the presence of a zero mode in the dispersion 
of the model at $k=0$ and $k=2\pi$ (cf. Eq.~\eqref{eq:ham-diag}). 
In~\ref{app:flatX} we derive analytically that this zero mode yields that
$\chi_A^x\simeq L$ for large $L$ (see Eq.~(\ref{eq:mu-nm})).
%
%
The same volume scaling $\simeq L$ is observed in short-range 
quantum spherical models that exhibit magnetic order~\cite{Wald20,Wald20-1,alba2021entanglement}.  
This reflects the fact that, although the dispersion of the model 
is dramatically affected by the long-range interactions,
the leading behavior of the magnetic susceptibility is dominated by the
zero mode, similar to the short-range case.
Now, let us decompose $\mathbb{X}_A$ as
\begin{equation}
	\label{eq:x-deco}
	\mathbb{X}_{A}=\chi_A^x|1\rangle\langle1|+\mathbb{X}'_A, 
\end{equation}
where $\chi_A^x$ is given in Eq.~\eqref{eq:su-def}, and $|1\rangle$ is the flat vector restricted 
to $A$. We exploit the fact that $\chi_A^x={\mathcal O}(L)$ and
consider the transposed correlation matrix\footnote{The transposition
does not affect the eigenvalues.}
$\mathbb{C}_A^T=\mathbb{P}_A\cdot \mathbb{X}_A$ 
(cf. Eq.~\eqref{eq:Corr-def}). By using Eq.~\eqref{eq:x-deco}, we obtain 
\begin{equation}
	\label{eq:xs}
	\mathbb{P}_A\cdot\mathbb{X}_A=\chi_A^x \mathbb{P}_A
	|1\rangle\langle1|+\mathbb{P}_A\cdot \mathbb{X}'_A.
\end{equation}
We can now neglect the second term in Eq.~\eqref{eq:xs} because it is subleading compared 
to the first one. Importantly, the matrix $\mathbb{P}_A\cdot\mathbb{X}_A$ is 
not hermitian. However, in the limit $L\to\infty$ it is 
easy to identify left and right eigenvectors, $\ket{u_R}$ and
$\bra{u_L}$ respectively, by inspection. They are given by
%
%
\begin{equation}
	\label{eq:ur}
	|u_R\rangle=\mathbb{P}_A|1\rangle,\quad \langle u_L|=\langle1|, 
\end{equation}
as can be seen by directly applying $\mathbb{C}_A^T$ to them, viz.,
\begin{align}
\mathbb{C}_A^T\ket{u_R} \simeq \langle1| \mathbb{P}_A|1\rangle \chi_A^x \mathbb{P}_A |1\rangle,
\quad
\bra{u_L} \mathbb{C}_A^T \simeq \chi_A^x \bra{1}\mathbb{P}_A\ket{1} \bra{1}.
\end{align}
As it is now clear from Eq.~\eqref{eq:xs}, the largest eigenvalue of $e_1$ of $\mathbb{C}_A$ is 
\begin{equation}
	\label{eq:e1}
	e_1 = \chi_A^x
	\langle1|\mathbb{P}_A|1\rangle=\bra{1}\mathbb{X}_A\ket{1}\bra{1} \mathbb{P}_A \ket{1}.
\end{equation}
We should mention that the same decomposition in Eq.~\eqref{eq:x-deco} was employed in Ref.~\cite{Botero04} 
to analyze the contribution of the zero mode to the ES in the harmonic chain. Moreover, 
the same decomposition 
has been employed to study the entanglement gap in the ordered phase of the two-dimensional quantum 
spherical model~\cite{Wald20-1,alba2021entanglement} (see also~\cite{Wald20}). 

Eq.~\eqref{eq:e1} shows that the finite-size scaling of the entanglement gap in the ferromagnetic 
phase is governed by the zero mode of the dispersion in Eq.~\eqref{eq:ham-diag}. Specifically, as it is 
clear from the lack of spatial structure of $|1\rangle$, $\chi_A^x$ is directly determined by the 
zero mode. On the other hand, the susceptibility $\chi_A^t$ is sensitive to the dispersion of the 
model. Crucially, both $\chi_A^x$ and $\chi_A^t$ can be determined analytically in the large $L$ limit. 
The derivation employs standard tools such as Poisson's summation formula and the Mellin transform, and 
it is reported in~\ref{app:mu}, ~\ref{app:flatX} and~\ref{app:pcorr}. The leading and first 
subleading contributions of $\chi_A^x$ in the large $L$ limit are 
\begin{equation}
	\label{eq:chix-an}
\chi_A^x\simeq
 \frac{1}{4}\sqrt{\frac{g}{2\mu}}
 +\frac{\sqrt{g}}{\pi}\sin\left(\frac{\pi }{4}\alpha\right)
 \Gamma\left(-1-\frac{\alpha}{2}\right)
 \left(  2^{1-\frac{\alpha}{2}} - 2^{3}\right)
 \zeta\left(-1-\frac{\alpha}{2}\right) L^{\frac{\alpha}{2}}, 
\end{equation}
where $\zeta(x)$ is the Riemann zeta function, and $\Gamma(x)$ is the Euler gamma function. 
The first term in Eq.~\eqref{eq:chix-an} is the zero-mode contribution, which is simply obtained 
by isolating the term with $k=0$ in Eq.~\eqref{eq:xx}. Since $\mu={\mathcal O}(L^{-2})$ in the ordered 
phase (see Fig.~\ref{fig:mu}), this term is ${\mathcal O}(L)$. The second term is ${\mathcal O}(L^{\alpha/2})$,  and it is subleading because $0<\alpha<2$. In Eq.~\eqref{eq:chix-an} we neglected 
$o(L^{\alpha/2})$ terms, which are reported in~\ref{app:flatX}. 
Eq.~\eqref{eq:chix-an} holds at the critical point as well, although it is not useful to 
determine the scaling of the entanglement gap since Eq.~\eqref{eq:e1chi} does not hold true at 
criticality. At the critical point one has $\mu={\mathcal O}(L^{-\alpha})$, which implies that 
both terms in Eq.~\eqref{eq:chix-an} are of the same order. 
It is important to stress that both at the critical point, as well as in the ordered phase, the 
terms  in Eq.~\eqref{eq:chix-an} depend only on the low-energy part 
of the dispersion of the QSM. In particular, the second term in Eq.~\eqref{eq:chix-an} does not depend on 
the cutoff $\Lambda$ introduced to regularize the behavior of the correlators. The second term in 
Eq.~\eqref{eq:chix-an} is one of an infinite number of terms that determine the universal behavior upon 
approaching the critical point. These terms are reported in~\ref{app:flatX}. 

Similarly, we obtain the leading behavior for $\chi_A^t$ as (see~\ref{app:pcorr})
\begin{equation}
	\label{eq:chip-an}
 \chi_A^t\simeq
 \frac{1}{\sqrt{g}} \frac{2}{\pi}
 \left(4- 2^{\frac{\alpha}{2}} \right) 
 \Gamma\left(\frac{\alpha}{2}-1\right)
 \sin\left(\frac{\pi}{4} \alpha \right)
 \zeta\left( \frac{\alpha}{2} -1 \right)  L^{-\frac{\alpha}{2}}
\end{equation}
Clearly, $\chi_A^t$ vanishes in the limit $L\to\infty$, in contrast to $\chi_A^x$ 
(cf. Eq.~\eqref{eq:chix-an}). 
Again, the behavior of $\chi_A^t$ is determined by the universal low-energy part of the 
dispersion of the model. Using Eqs.~\eqref{eq:e1},~\eqref{eq:chix-an} and~\eqref{eq:chip-an}, 
we obtain 
\begin{equation}
	\label{eq:e1-scaling}
	e_1\simeq C'_\alpha L^{1-\frac{\alpha}{2}}= 
	\frac{1}{\pi} 
	\left(\frac{1}{\sqrt{g}}-\frac{1}{\sqrt{g}} \right) \left(4- 2^{\frac{\alpha}{2}} \right) 
 \Gamma\left(\frac{\alpha}{2}-1\right)
 \sin\left(\frac{\pi}{4} \alpha \right)
 \zeta\left( \frac{\alpha}{2} -1 \right)  L^{1-\frac{\alpha}{2}}. 
\end{equation}
As it is clear from Eq.~\eqref{eq:e1-scaling} the eigenvalue $e_1$ diverges in the 
limit $L\to\infty$ because $0<\alpha<2$. Moreover, the constant $C'_\alpha$ 
depends on the low-energy properties of the QSM. 
Finally, we obtain that the entanglement gap $\delta\xi$ in the large $L$ limit vanishes as 
\begin{equation}
	\label{eq:dxi-final}
	\delta\xi\simeq C_\alpha L^{-\frac{1}{2}+\frac{\alpha}{4}}, \quad\mathrm{with}\,\,
	C_\alpha=\frac{1}{\sqrt{C'_\alpha}}, 
\end{equation}
with $C'_\alpha$ as defined in Eq.~\eqref{eq:e1-scaling}. 

It is interesting to compare the result in Eq.~\eqref{eq:e1-scaling} with the 
scaling of the entanglement gap in the magnetically ordered phase of the 
two-dimensional QSM~\cite{alba2021entanglement}. 
Similar to Eq.~\eqref{eq:dxi-final}, $\delta\xi$ exhibits  a power-law decay with $L$. 
Precisely, for the $2D$ QSM one has the behavior~\cite{alba2021entanglement}  
\begin{equation}
	\label{eq:dxi-2d}
	\delta\xi\simeq \frac{\Omega}{\sqrt{L\ln(L)}}, 
\end{equation}
where $\Omega$ is a constant that depends on the geometry of the bipartition and 
on the low-energy properties of the QSM. In particular $\Omega$ is dramatically affected 
by the presence of corners in the boundary between $A$ and the rest. 
Notice that the multiplicative logarithmic correction in Eq.~\eqref{eq:dxi-2d}, which reflects a multiplicative 
logarithmic correction in $e_1$, is a genuine consequence of the model being two-dimensional, 
and it is absent in the $1D$ long-range QSM. 

Finally, it is interesting to observe that on the critical line (see Fig.~\ref{fig:pd}) 
one has that $\mu={\mathcal O}(L^{-\alpha})$. Thus, by using Eq.~\eqref{eq:e1-scaling} 
one obtains that $e_1\simeq{\mathcal O}(1)$. However, this is not accurate because we numerically 
observe that at criticality $e_1$ diverges, although slowly, signaling that the entanglement gap 
vanishes at criticality as well. This is somewhat similar in the $2D$ 
QSM~\cite{Wald20-1}, where the same approximation from Eq.~\eqref{eq:x-deco} leads 
to an inaccurate scaling for the entanglement gap. The reason is that at the critical point the 
eigenvector of $\mathbb{X}$ exhibits a non trivial structure, i.e., it is different from the 
flat vector $|1\rangle$.

\section{Numerical benchmarks}
\label{sec:numerics}
%
%
\begin{figure}[t]
\centering
\includegraphics[width=.495\textwidth]{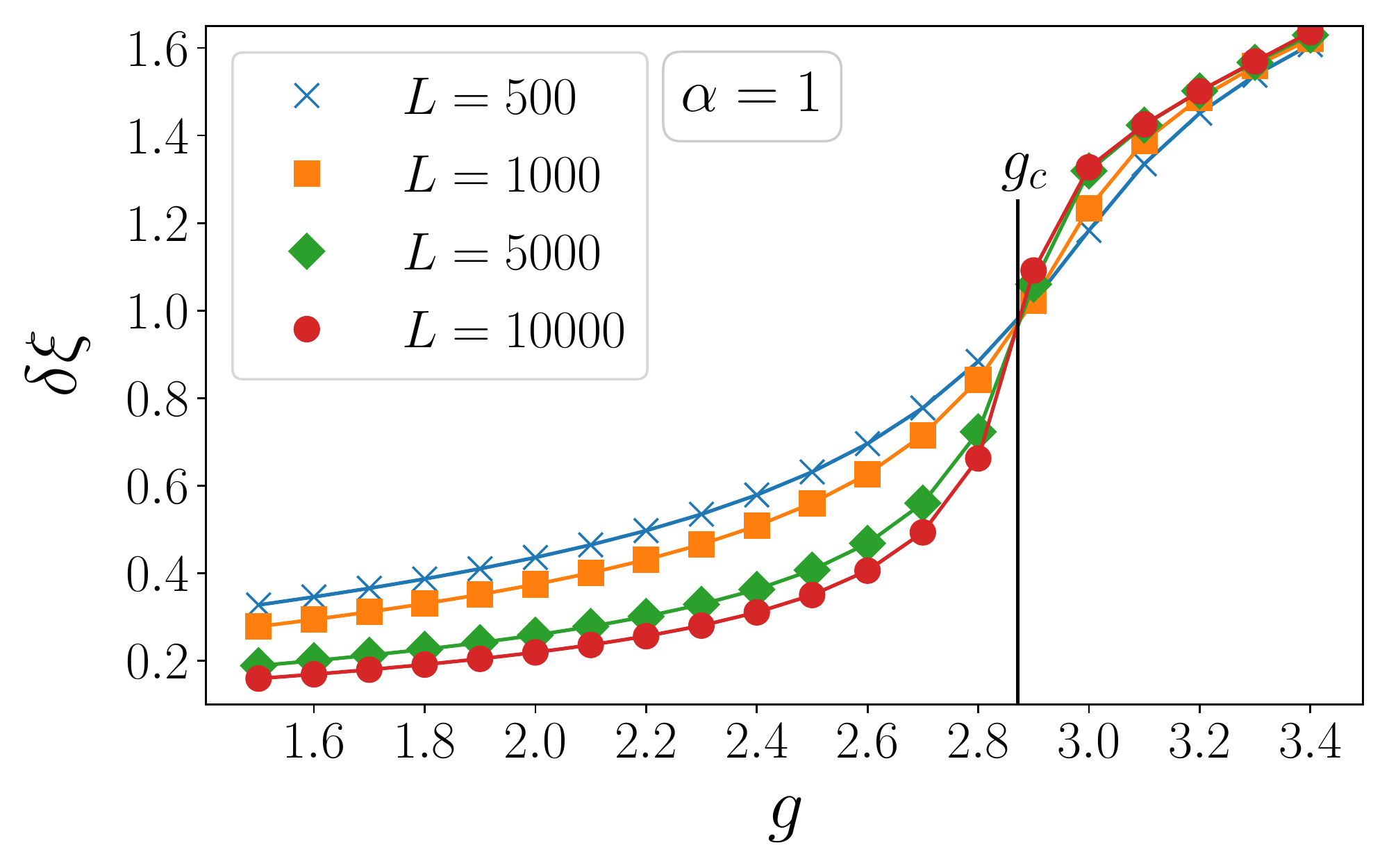}
\includegraphics[width=.495\textwidth]{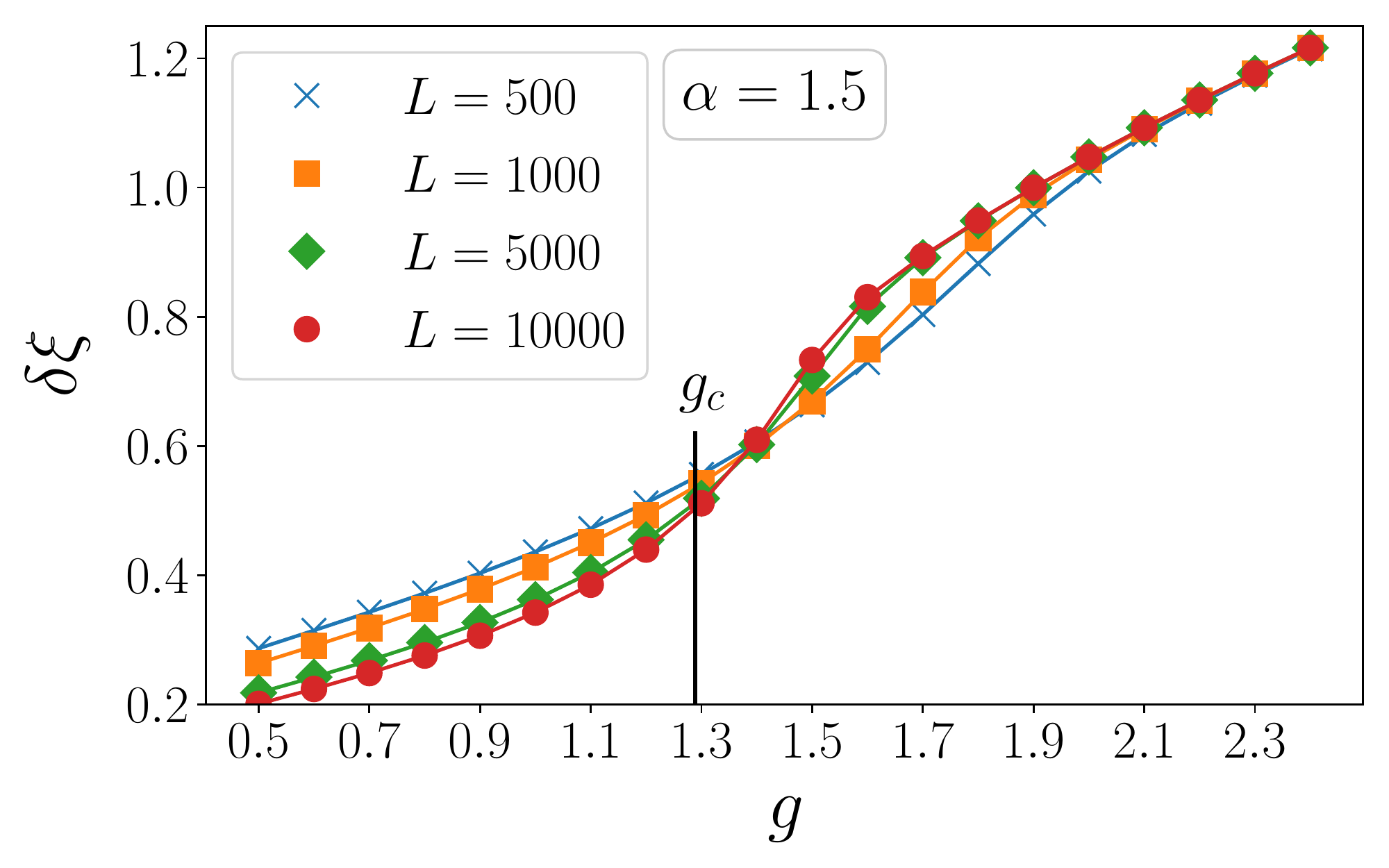}
\caption{ Lowest entanglement gap $\delta\xi$ in the ground-state ES of the QSM 
 with long-range interactions. Here we consider the half-chain ES (see Fig.~\ref{fig:cartoon}), 
 plotting $\delta\xi$ 
 versus the coupling $g$. The left and right panels correspond to $\alpha=1$ and $\alpha=1.5$, 
 respectively. The different symbols are for different system size $L$. 
 The vertical lines mark the quantum critical point at $g_c$. In the 
 paramagnetic phase for $g>g_c$, $\delta\xi$ attains a finite value in the limit $L\to\infty$. 
 For $g\le g_c$ the entanglement gap $\delta\xi$ vanishes in the limit $L\to\infty$. 
}
\label{fig:e-gap}
\end{figure}
\begin{figure}[t]
\includegraphics[width=.495\textwidth]{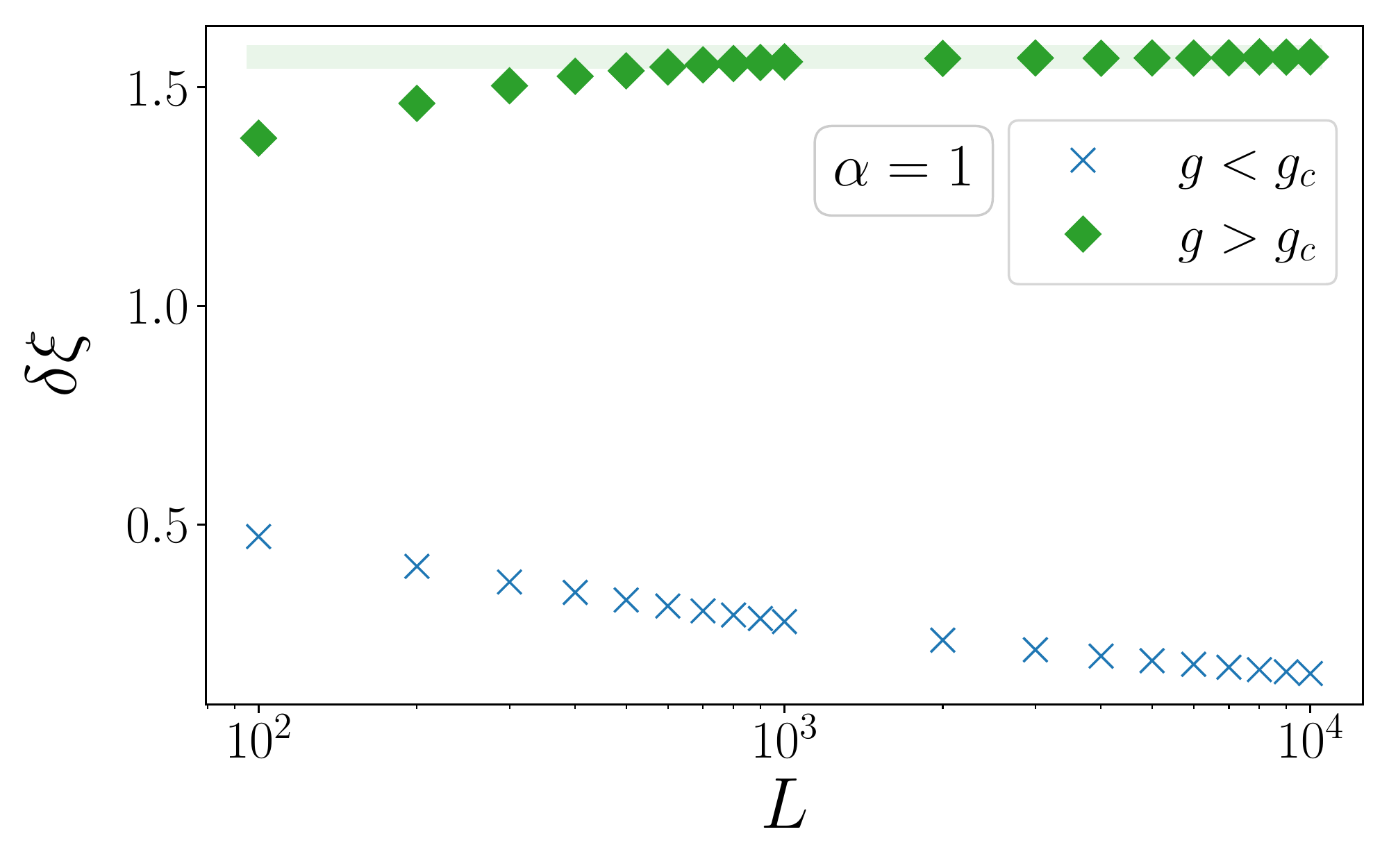}
\includegraphics[width=.495\textwidth]{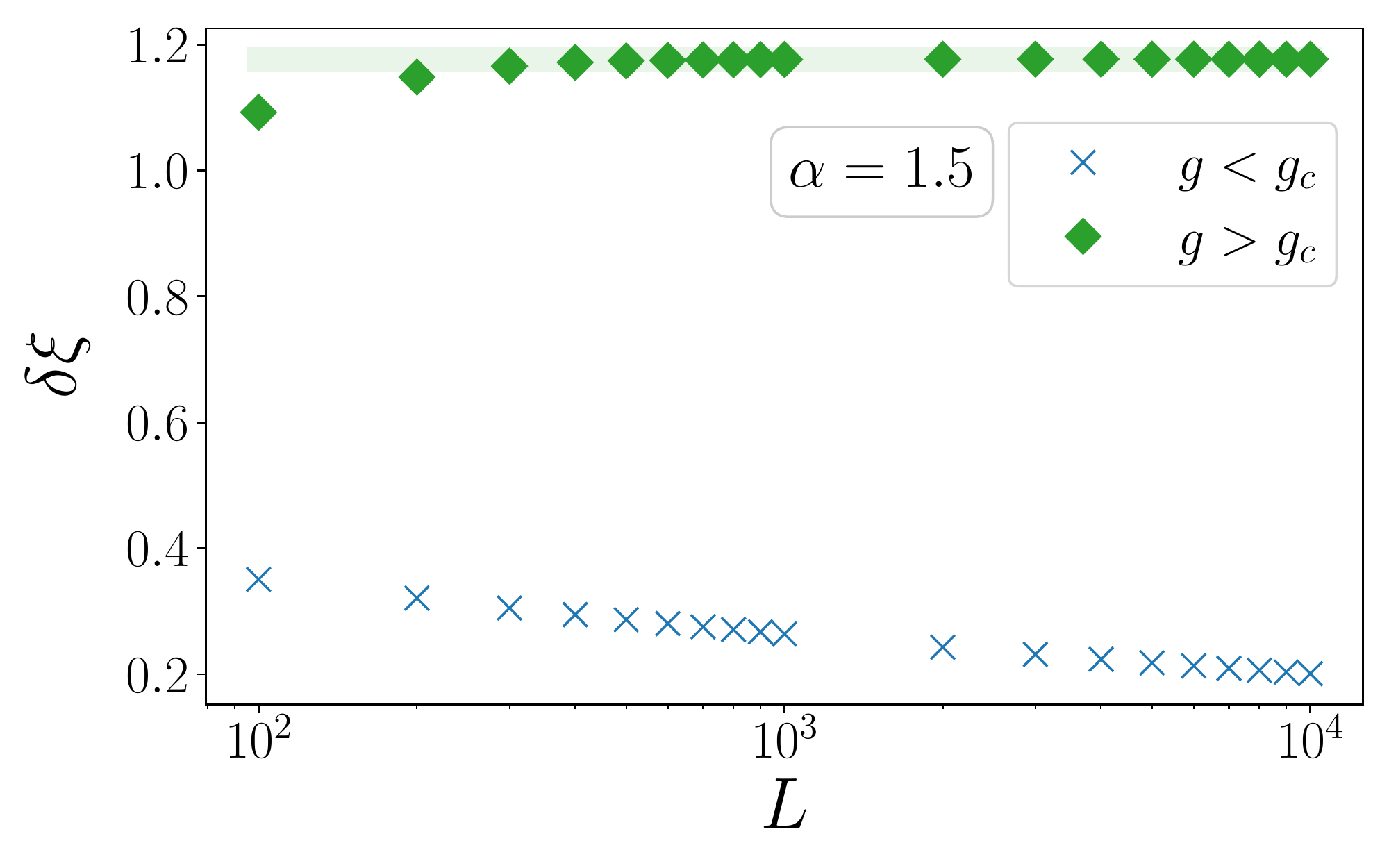}
\caption{{Qualitative behavior of the entanglement gap.}
Exemplary data from Fig.~\ref{fig:e-gap} for the entanglement gap $\delta\xi$ are shown 
in the ferromagnetic and in the paramagnetic phase. In the ferromagnetic phase
the gap vanishes, albeit slowly. Conversely, in the paramagnetic phase the gap remains finite.}
\label{fig:xi_quali}
\end{figure}

%
Here we provide numerical benchmarks of the results of section~\ref{sec:main}. 
We start discussing the general structure of the entanglement gap across the phase diagram 
of the QSM (see Fig.~\ref{fig:pd}). In Fig.~\ref{fig:e-gap} 
we show the entanglement gap $\delta\xi$ as a function of the quantum coupling $g$ across 
the phase transition. The data are obtained by computing the correlation functions 
in Eq.~\eqref{eq:Corr-def} 
with the spherical parameter $\mu$ obtained by numerically solving Eq.~\eqref{eq:constraint}, and 
by using Eq.~\eqref{eq:ES-gap}. The left and right panel show results for $\alpha=1$ and $\alpha=3/2$, 
respectively. The different symbols correspond to different system sizes $500\le L\le 10000$. 
In Fig.~\ref{fig:e-gap} we consider the bipartition with $L_A=L/2$ (see Fig.~\ref{fig:cartoon}). 
The vertical lines in 
Fig.~\ref{fig:e-gap} mark the critical coupling $g_c$.
For $g>g_c$ the entanglement
gap $\delta\xi$ attains a finite value in the limit $L\to\infty$ as 
can be seen from Fig.~\ref{fig:xi_quali}.

On the other hand, in the 
ordered phase for $g<g_c$ the data suggests a vanishing $\delta\xi$ in the limit $L\to\infty$, 
although sizeable finite $L$ effects are visible. 
%
%
\begin{figure}[t]
\centering
\includegraphics[width = .6\columnwidth]{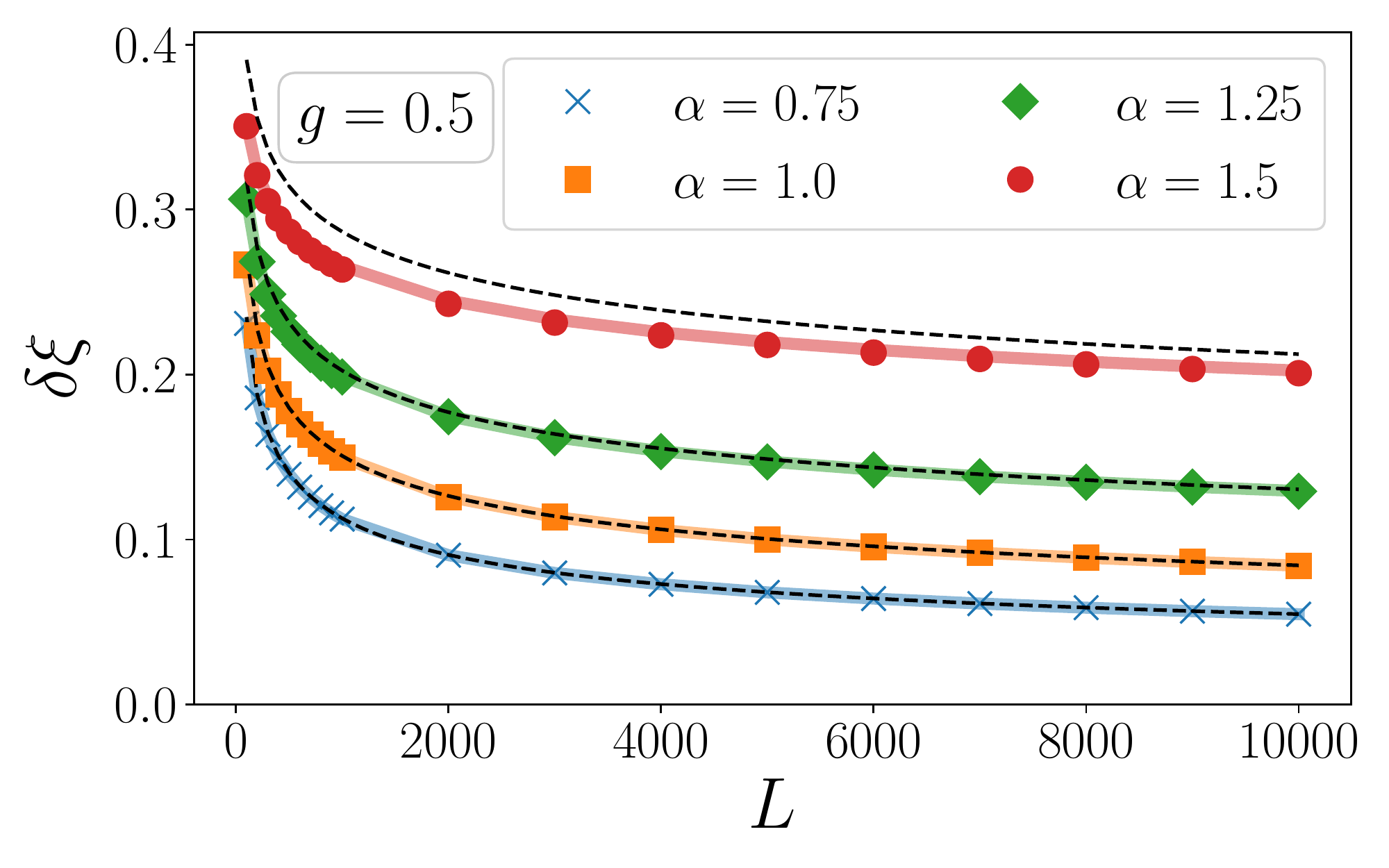}
\caption{ Finite-size scaling of the lowest entanglement gap $\delta\xi$ in 
 the ferromagnetic phase of the long-range QSM. We plot 
 $\delta\xi$ versus $L$ at fixed $g=1/2$. The results are for the half-system 
 ES (see Fig.~\ref{fig:cartoon}). Different symbols correspond to 
 different values of the long-range exponent $\alpha$. The continuous lines are obtained 
 by using~\eqref{eq:e1chi}. The dash-dotted lines are obtained from the analytic 
 results~\eqref{eq:e1-scaling} in the large $L$ limit. 
}
\label{fig:dxi-chi}
\end{figure}
%
%
The finite-size scaling of $\delta\xi$ is  investigated in Fig.~\ref{fig:dxi-chi} plotting 
$\delta\xi$ versus $L$ for fixed 
$g=1/2$, i.e., in the ferromagnetic phase. The different symbols denotes results for different 
values of the long-range exponent $\alpha$. For all the values of $\alpha$ considered, $\delta\xi$ 
exhibits vanishing behavior in the limit $L\to\infty$. The continuous line in Fig.~\ref{fig:dxi-chi} is the prediction obtained by numerically computing $\chi_A^x$ and $\chi_A^t$ (cf. Eq.~\eqref{eq:su-def}), and by 
employing~\eqref{eq:e1-scaling}. The agreement between 
the lattice results and the analytic results in the asymptotic limit $L\to\infty$ is 
perfect. Finally, the dash-dotted line in Fig.~\ref{fig:dxi-chi} is Eq.~\eqref{eq:dxi-final}. 
The data are in perfect  agreement with~\eqref{eq:dxi-final}, except for $\alpha=1.5$,  for which  
some deviations are visible. These are attributed to the finite $L$. Indeed, similar deviations 
are also visible for $\mu$ in Fig.~\ref{fig:mu}, where we show much larger system sizes up to 
$L\approx 10^6$. 

\section{Conclusions}
\label{sec:concl} 

We characterized the finite-size scaling  of the entanglement gap in the long-range 
$1D$ quantum spherical model. Our main result 
is given by Eq.~\eqref{eq:dxi-final}. We showed  that in the ferromagnetically ordered phase 
of the long-range QSM the entanglement gap vanishes in the thermodynamic limit as $\simeq C_\alpha 
L^{-1/2+\alpha/4}$.  The prefactor $C_\alpha$ of the decay depends only on the low-energy properties of 
the model. This behavior is different from the $2D$ quantum spherical model, where 
the power-law decay of the entanglement gap is accompanied by multiplicative 
logarithmic corrections~\cite{alba2021entanglement}. 

Let us now mention some possible future directions. First, it would be interesting to 
determine the finite-size scaling of the entanglement gap on the critical line as a function 
of the long-range exponent $\alpha$. This is in general a challenging task because Eq.~\eqref{eq:e1chi} 
is not valid at criticality. 
An interesting question is whether it is possible to 
determine the behavior of the distribution of the ES 
levels~\cite{calabrese2008entanglement}, and how it is affected by 
the long-range interactions. The main challenge is that Conformal Field Theory does not 
hold in the presence of long range interactions. One of our main results is 
Eq.~\eqref{eq:e1}, which  confirms that there is a robust relationship 
between the entanglement gap and standard witnesses of magnetic order, such 
as $\chi_A^x$ and $\chi_A^t$. It would be important to understand whether Eq.~\eqref{eq:e1} 
survives for the $O(N)$ models away from the $N\to\infty$ limit. 
It would be also interesting to investigate the effects of disorder on entanglement properties 
of the long-range QSM, by using the replica trick to perform disorder averages~\cite{kosterlitz1976spherical,hornreich1982thermodynamic,jagannathan1989the,vojta1993spherical,vojta1996critical}. 
Another important research direction is to investigate entanglement scaling after quantum 
quenches in the long-range QSM, using the results of Refs.~\cite{chandran2013equilibration,barbier2019pre,barbier2022generalized,henkel2022quantum}. Finally, it would be interesting to investigate 
the negativity spectrum~\cite{ruggiero-2016,shapourian-2019,xhek-2020_A} in the long-range QSM.

\section*{Acknowledgement}
The authors are  grateful to M.~Henkel for useful discussions that helped
advance this project.

\appendix

\section{Critical coupling $g_c(\alpha)$}
\label{app:gc}

\noindent
Here we derive for generic $\alpha$ the critical coupling $g_c$ of the 
second order phase transition that divides the paramagnetic phase for $g>g_c$ 
from the ordered phase at $g<g_c$ (see Fig.~\ref{fig:pd}). 

Let us start with the two-point auto-correlation function~\cite{Wald20}
\begin{equation}
	\label{eq:xnn}
	\mathbb{X}_{nn} = \frac{g}{2L} \sum_{k\in\mathcal{B}} \frac{1}{E_k}.
\end{equation}
The spherical constraint, Eq.~\eqref{eq:gc}, in the thermodynamic limit $L\to\infty$ reads
\begin{equation}
	\label{eq:const-1}
	1=\int_{0}^{2\pi} \frac{\D k}{2\pi}
	\frac{\sqrt{g}/2}{\sqrt{2\mu+ \left(2(1-\cos k)\right)^{\frac{\alpha}{2}}}}.
\end{equation}
In order to extract $g_c(\alpha)$ we directly integrate the spherical
constraint for $\mu=0$ and find
\begin{equation}
\frac{2}{g_c}=\int_{0}^{2\pi}\frac{dk}{2\pi}\frac{1}{E_k}=\frac{2^{-\frac{\alpha
}{2}}\Gamma\left(1/2-\alpha/4\right)}{\sqrt{g_c\pi}\Gamma\left(1-\alpha/4\right)
}.
\end{equation}
Thus, we obtain 
\begin{equation}
g_c=
2^{\alpha+2}\pi
\left(\frac{\Gamma\left(1-\alpha/4\right)}{
\Gamma\left(1/2-\alpha/4\right)}\right)^{2}.
\label{eq:gc-app}
\end{equation}
The behavior of $g_c$ as a function of $\alpha$ is reported in Fig.~\ref{fig:pd}. 
Notice that we integrated over the full Brillouin zone to obtain $g_c$, which reflects 
that $g_c$ is non universal.

\section{Finite-size scaling of the spherical parameter}
\label{app:mu}

\noindent
Let us now extract the finite-size scaling (FSS) of the spherical parameter $\mu$, which 
is determined by solving 
\begin{equation}
\label{eq:mu-fs-app}
\frac{2}{\sqrt{g}} = \frac{1}{L}\sum_{n=0}^{L-1}
\frac{1}{\sqrt{2\mu + (2(1-\cos\left(2\pi n/L\right)))^{\alpha/2}}}.
\end{equation}
The strategy is to use {\it Poisson's summation formula} 
\begin{equation}
\label{eq:poisson}
 \sum_{n=a}^b f(n) = \frac{f(a)+f(b)}{2} + \int_a^b f(x) {\D}x
 +2\sum_{p=1}^\infty \int_a^b f(x) \cos(2\pi p x ) {\D} x
\end{equation}
to split~\eqref{eq:mu-fs-app} into a thermodynamic contribution 
\footnote{Although this contribution is formally equivalent to the
thermodynamic contribution, the spherical parameter $\mu$ is
still finite-size dependent.} 
and a finite-size one. 
It is useful to observe that in our case (cf.~\eqref{eq:mu-fs-app}) $a=0$ and $b=L-1$ and that
$f(0)=f(L)$. Thus, it is convenient to add and subtract in~\eqref{eq:poisson}
the term with $n=b+1$. This allows us to get rid of the boundary contribution
in the right-hand-side of~\eqref{eq:poisson}. This means that we can use the
modified version of the Poisson summation formula as
\begin{equation}
\label{eq:poisson-1}
\sum_{n=a}^{b} f(n) =  \int_a^{b+1} f(x) {\D}x
+2\sum_{p=1}^\infty \int_a^{b+1} f(x) \cos(2\pi p x ) {\D} x, \quad\mathrm{if}\,\,f(a)=f(b+1).
\end{equation}
By using~\eqref{eq:poisson-1} we can rewrite~\eqref{eq:mu-fs-app} as 
\begin{multline}
\label{eq:step-1}
\frac{2}{\sqrt{g}} =
\frac{1}{L}\int_0^{L} \frac{\D x}{\sqrt{2\mu + \left(2(1-\cos(2\pi
x/L))\right)^{\alpha/2}}}\\
+ \frac{2}{L}\sum_{n=1}^{\infty} \int_0^{L}
	\frac{\cos\left(2\pi n x \right)\D x }{\sqrt{2\mu + \left(2(1-\cos(2\pi x/L))\right)^{\alpha/2}}}
\end{multline}
For the remainder of this section, we work in the long wavelength
approximation\footnote{In Ref.~\cite{Bran88}
it has been shown that this approximation recovers the dominant
FSS behavior of the model.} in which we expand $\cos(k)\approx 1-k^2/2$ in 
the denominators in~\eqref{eq:step-1}. This approximation affects the behavior of 
nonuniversal quantities at the transition, such as the value of the critical 
coupling. In the ferromagnetic phase the  long-wavelength 
approximation affects quantities that depend on the full dispersion 
of the model. However, as we are going to verify, the behavior of the 
entanglement gap is sensitive only to the lower-energy properties of the 
dispersion. This means that the results that we are going to derive apply 
to the model with the cosine dispersion as well. 

In the long-wavelength approximation, we can rewrite~\eqref{eq:step-1} as
\begin{equation}
\label{eq:step-2}
\frac{1}{\sqrt{g}} =
\int_0^{\Lambda} \frac{\D k}{2\pi}\frac{1}{\sqrt{2\mu + k^{\alpha}}}
+ 2\sum_{n=1}^{\infty} \int_0^{\Lambda} \frac{\D k}{2\pi}
\frac{\cos\left(n k L \right) }{\sqrt{2\mu + k^{\alpha}}}. 
\end{equation}
Here after applying the long wavelength approximation we multiplied the 
right-hand-size by two to account for the  fact that the two singularities at $k=0$ and $k=2\pi$ 
in the original dispersion contribute equally. 
Here we  also extend the Brillouin zone from $[0,2\pi]\to[0,\Lambda)$, 
introducing the ultraviolet cutoff $\Lambda$. 
To proceed, we need to extract the large $L$ behavior of  the two terms in~\eqref{eq:step-2}. 
The integral in the first term in~\eqref{eq:step-2} is readily evaluated as in section~\ref{app:gc}. We find
\begin{equation}
	\label{eq:step-3}
	\int_0^\Lambda \frac{\D k}{2\pi} \frac{1}{\sqrt{2\mu+k^{\alpha}}}
\stackrel{\mu\to0}{\simeq}\frac{2}{\sqrt{g_c}}+\frac{\Gamma\left(\frac{1}{2}
-\frac{1}{\alpha} \right)\Gamma\left(1+\frac{ 1}{\alpha}\right) }{
2\pi^{3/2} }
\left(2\mu\right)^{\frac{1}{\alpha}-\frac{1}{2}}. 
\end{equation}
Here we considered the limit $\mu\to0$ because we are interested in the 
magnetically ordered phase and in the critical point, where $\mu=0$ in the 
thermodynamic limit $L\to\infty$. 
In~\eqref{eq:step-3}  we identified the critical coupling $g_c$  as
$ g_c = 4\pi^2 (2-\alpha)^2/\Lambda^{2-\alpha} $. Notice that 
$g_c$ depends on the cutoff $\Lambda$, as expected because it is a nonuniversal
quantity. On the other hand, the second term in~\eqref{eq:step-3} does not depend on 
$\Lambda$. We also checked that higher orders in the expansion in 
the limit $\mu\to0$ would depend on the cutoff $\Lambda$. The leading order in
$\mu$ reveals the onset of mean-field for $\alpha \leq 2/3$.

The analysis of the second term on the right-hand side in~\eqref{eq:step-2} 
is more involved and can be performed by employing the Mellin transform~\cite{NIST:DLMF}. 
To proceed, we first define the function $f(n)$ as 
\begin{equation}
	\label{eq:fn-def}
f(n) := \int_0^\Lambda \frac{\D k}{2\pi} \frac{\cos(k L n)}{\sqrt{2\mu +
k^{\alpha}}}, 
\end{equation}
and analyze  the series $\sum_{n=1}^\infty f(n)$ (cf.~\eqref{eq:step-2}) 
by using standard regularization techniques~\cite{contino-2002}.
The Mellin transform $\hat g(s)$ of a function $g(x)$ is defined as
\begin{equation}
	\label{eq:mellin-def}
	\hat g(s) = \int_0^\infty\D x \, g(x) x^{s-1}. 
\end{equation}
The inverse of the Mellin transform is performed as 
\begin{equation}
	\label{eq:iMf}
	g(x)=\frac{1}{2\pi \II}\int_{c-\II\infty}^{c+\II\infty}\D s\, x^{-s}\hat g(s), 
\end{equation}
where $c$ is chosen in the so-called fundamental strip. 

For the function $f(n)$ (cf.~\eqref{eq:fn-def}) we obtain in 
the limit $\mu\to0$ 
\begin{equation}
\hat{f}(s)
\simeq \frac{(2\mu)^{\frac{2-\alpha-2s}{2\alpha}}}{2\pi^{3/2}\alpha L^s}
\Gamma\left(\frac{1}{2}+\frac{s-1}{\alpha}\right)
\Gamma\left(\frac{1-s}{\alpha}\right)
\cos\left(\frac{\pi}{2}s\right)\Gamma(s).
\label{eq:Mf}
\end{equation}
Again, in the expansion around $\mu=0$ in~\eqref{eq:Mf}, we neglect all 
the higher-order terms that depend on the cutoff $\Lambda$. 
The condition that the integral over $k$ in~\eqref{eq:fn-def} is defined 
for $k\to0$ implies that $\mathrm{Re}(s)<1$. On the other hand, the condition 
that the integral is well-defined at $\Lambda\to\infty$ 
implies that $\mathrm{Re}(s)>1-\alpha/2$. As we have a finite cutoff $\Lambda$ and 
we are not interested in cutoff-dependent contributions, we have the condition $\mathrm{Re}(s)<1$. 
Importantly, as it is clear from~\eqref{eq:Mf} we can extend the fundamental strip beyond 
$s=1$ because the cosine function removes the simple pole of $\Gamma((1-s)/\alpha)$ at $s=1$. 
We can now write the series $\sum_{n=1}^\infty f(n)$ (cf.~\eqref{eq:step-2}) as
\begin{equation}
	\label{eq:mell-inv}
	\sum_{n=1}^\infty f(n) =
	\frac{1}{2\pi \II }\int_{c-\II\infty}^{c+\II\infty}\D s\, \hat{f}(s)
	\sum_{n=1}^\infty n^{-s}
=
	\frac{1}{2\pi \II }\int_{c-\II\infty}^{c+\II\infty}\D s\, \hat{f}(s)
	\zeta(s). 
\end{equation}
Here we used the definition of the Riemann zeta function $\zeta(s)$, and we 
have $\mathrm{Re}(c)>1$.  
%
\begin{figure}[t]
\centering
\includegraphics[width=.4\columnwidth]{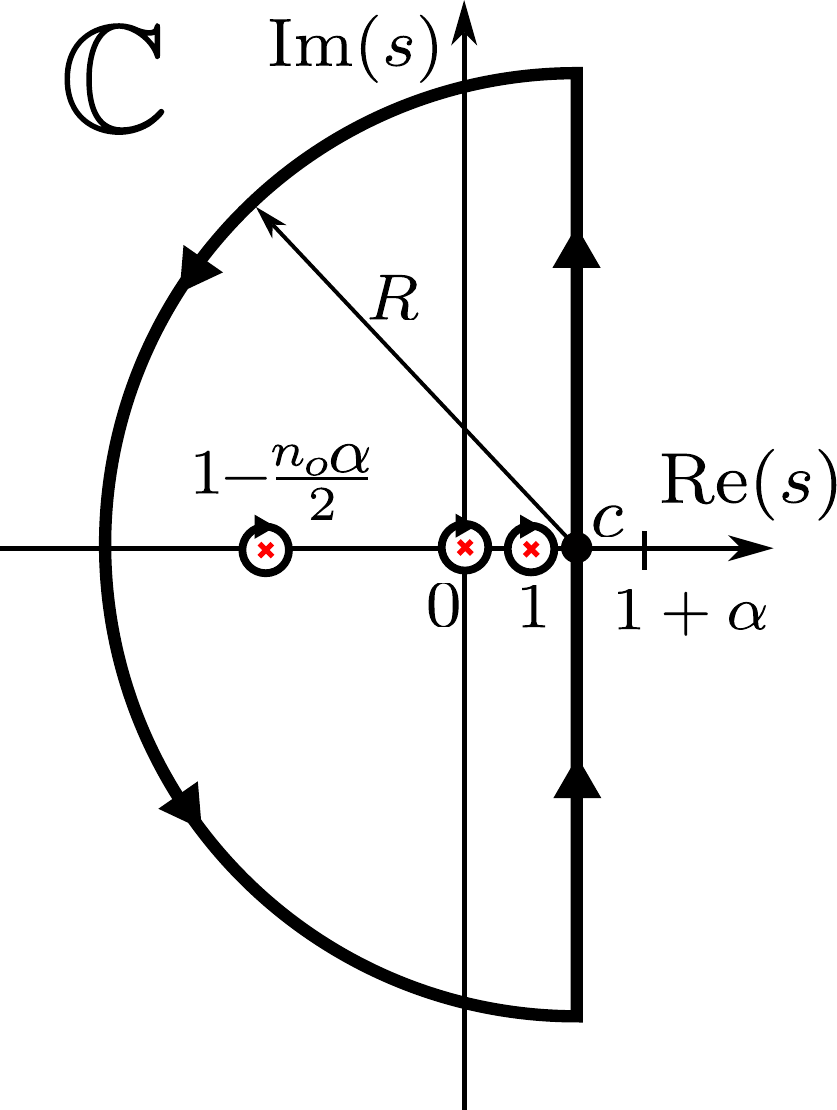}
\caption{Integration contour in the complex plane $\mathrm{Im}(s)$ versus $\mathrm{Re}(s)$ 
 used to compute the inverse Mellin transform in~\eqref{eq:mell-inv}. The vertical part 
 of the contour corresponds to fixed $\mathrm{Re}(s)=c$, with $1<c<1+\alpha$, where the 
 integrand in~\eqref{eq:mell-inv} is analytic. The crosses are the poles of the integrand. 
 The simple pole at $s=1$ is due to the Riemann zeta function in~\eqref{eq:mell-inv}. The poles 
 at $s=0$ and at $s=1-n_o\alpha/2$ are due to the functions $\Gamma(s)$ and $\Gamma(1/2-(s-1)/\alpha)$ 
 in~\eqref{eq:Mf}. Here $n_o:=2p+1$ with 
 $p\in\mathbb{N}$. The remaining poles of~\eqref{eq:mell-inv} 
 are removed by $\zeta(s)$ and by $\cos(\pi s/2)$. 
}
\label{fig:contour}
\end{figure}
%
Notice that the fact that the integrand in~\eqref{eq:mell-inv} is analytic 
for $1<\mathrm{Re}(s)<1+\alpha$ ensures that it is possible to define the 
fundamental strip for $s>1$. To proceed, we perform the integral over $s$ in~\eqref{eq:mell-inv} 
in the complex plane. 
To choose the suitable contour we observe that the spherical parameter decays algebraically
with increasing $L$, both at the critical point and in the ordered phase.
This suggests the finite-size scaling behavior of $\mu$ as $\mu\propto L^{-\sigma}$ with
$\sigma>0$. By using~\eqref{eq:Mf}, this suggests the scaling of $\hat f(s)$ as
\begin{equation}
	\label{eq:hat-f-scal}
	\hat{f}(s) \propto
	L^{s(\sigma/\alpha-1)} L^{\sigma (\alpha-2)/(2\alpha)}. 
\end{equation}
Since $\alpha<2$,  the second term in~\eqref{eq:hat-f-scal} always 
decays for $L\to\infty$, whereas the behavior of the
first one is different for $\sigma\geq \alpha$ and for $\sigma<\alpha$.
However, we can exclude that $\sigma<\alpha$ because for $\alpha\to0$, i.e., for
the infinite-range model, this would yield a finite $\mu$.
Hence, we consider $\sigma\geq \alpha$.
Thus, a consistent finite-size analysis suggests to close the complex
contour at $\operatorname{Re}(s) \to-\infty$, as shown in Fig.~\ref{fig:contour}. 
The integral is determined by the singularities within the contour, which we now 
discuss. 

First, the Riemann zeta function $\zeta(s)$ 
has a simple pole at $s=1$. The gamma function $\Gamma(s)$ has poles at  $s=-n$ with $n\in\mathbb{N}$ an 
integer. The function $\Gamma((1-s)/\alpha)$ has poles at $s=n\alpha+1$, with
$n\in\mathbb{N}/\{0\}$, and at $s=1-(2n+1)\alpha/2$, with $n\in\mathbb{N}$. 
Notice that the poles at $1+n\alpha$ are not within the integration contour (see Fig.~\ref{fig:contour}), 
and we can neglect them. Moreover, the poles at $s=-n_o$ with
$n_o$ odd positive integers cancel out with the term $\cos(\pi/2 s)$ in~\eqref{eq:mell-inv}.
On the other hand, the poles at $s=-n_e$ with $n_e$ an arbitrary positive 
even integer do not contribute because $\zeta(-n_e)=0$.  
In conclusion, the only poles $s^*$ that contribute to the integral in~\eqref{eq:mell-inv} are 
\begin{equation}
	\label{eq:table}
	s^*=\left\{
		\begin{array}{cc}
			0 & \\
			1 & \\
			1-\frac{(2p+1)\alpha}{2} & p\in\mathbb{N}
		\end{array}
	\right.
\end{equation}
Thus, since the contribution of the circle in the contour in Fig.~\ref{fig:contour} vanishes 
for $R\to\infty$, from~\eqref{eq:mell-inv} we obtain that 
\begin{equation}
	\sum_{n=1}^\infty f(n)=\sum_{\mathrm{poles}\,s^*}\mathrm{Res}(\hat f(s)\zeta(s),s^*), 
\end{equation}
where $s^*$ are given in~\eqref{eq:table}. 
Specifically, the pole at $s=1$ gives the contribution 
\begin{equation}
	\label{eq:t11}
	\operatorname{Res}(\hat{f}(s)\zeta(s),s=1)
	= \frac{(2\mu)^{-\frac{1}{2}} }{4L},
\end{equation}
where we used that the residue of $\zeta(s)$ at $s=1$ is one. 
To proceed, we observe that  the singularities of $\Gamma(s)$ at $s=-p$ with $p$ an integer 
are simple poles,  with residue 
\begin{equation}
	\mathrm{Res}(\Gamma(s),-p)=\frac{(-1)^p}{p!}.
\end{equation}
This allows us to obtain the contribution at $s^*=0$ (cf.~\eqref{eq:table}) as 
\begin{equation}
\label{eq:t2}
\mathrm{Res}(\hat f(s)\zeta(s),0)=\mu^{-\frac{1}{2}+\frac{1}{\alpha}}r',\quad\mathrm{with}\,\,
r':=-\frac{2^{-\frac{5}{2}+\frac{1}{\alpha}}}{\pi^{\frac{3}{2}}}
\Gamma\left(\frac{1}{2}-\frac{1}{\alpha}\right)\Gamma\left(1+\frac{1}{\alpha}\right).
\end{equation}
Finally, let us consider the poles at $s=1-(2 p+1)\alpha/2$.
We obtain that
\begin{equation}
	\label{eq:t3}
	\mathrm{Res}\left(\hat f(s)\zeta(s),1-\frac{2p+1}{2}\alpha\right)=
	\mu^p  L^{\alpha  \left(p+\frac{1}{2}\right)-1} r_p,
\end{equation}
with $r_p$ defined as
\begin{equation}
\label{eq:t3}
	r_p:=
	\frac{(-1)^p 2^{p-1}}{\pi ^{\frac{3}{2}}
	p!} \Gamma \left(p+\frac{1}{2}\right)
  \sin \left(\frac{1}{4} \pi
   \alpha  (2 p+1)\right) \Gamma \left(-p \alpha -\frac{\alpha
   }{2}+1\right) \zeta \left(1-\frac{1}{2} (2 p+1) \alpha \right).
\end{equation}
Finally, putting together~\eqref{eq:t11}~\eqref{eq:t2} and~\eqref{eq:t3} we obtain 
\begin{equation}
	\label{eq:t4}
	\sum_{n=1}^\infty f(n)=\frac{(2\mu)^{-\frac{1}{2}}}{4L}+
	\mu^{-\frac{1}{2}+\frac{1}{\alpha}}r'+
	\sum_{p=0}^\infty \mu^p  L^{\alpha  \left(p+\frac{1}{2}\right)-1} r_p.
\end{equation}
Now, it is important to notice that at the critical point we expect
$\mu\propto L^{-\alpha}$. This implies that all the three contributions
in~\eqref{eq:t4} are of the same
order $L^{\alpha/2-1}$. Oppositely, in the ferromagnetically ordered phase 
one has $\mu\propto L^{-2}$, implying that in the large $L$ limit
the first term in~\eqref{eq:t4} is the leading one, whereas the other ones
are suppressed. Thus, to obtain the leading behavior of $\mu$ for $g<g_c$ 
it is sufficient to replace~\eqref{eq:mu-fs-app} with the equation
\begin{equation}
	\frac{1}{\sqrt{g}}-\frac{1}{\sqrt{g_c}}\simeq
    \frac{(2\mu)^{-\frac{1}{2}}}{2L}, 
\end{equation}
which allows us to readily find 
\begin{equation}
\mu=\frac{1}{8}\left(\frac{1}{\sqrt{g}}-\frac{1}{\sqrt{g_c}}\right)^{
-2 }\frac{1}{L^2}+o(L^{-2}),\quad\mathrm{for}\,\,g<g_c. 
\end{equation}
In particular, deep in the ferromagnetic phase, we find
\begin{equation}
	\mu \simeq\frac{g}{8} \frac{1}{L^2}.
\end{equation}
To extract the finite-size scaling of $\mu$ at the critical point, let us define 
$\gamma_\alpha$ as 
\begin{equation}
	\label{eq:mu-ansatz}
	\mu=\frac{\gamma_\alpha}{L^\alpha}. 
\end{equation}
After substituting the ansatz~\eqref{eq:mu-ansatz} in the gap equation~\eqref{eq:step-2} 
and setting $g=g_c$, we obtain the equation for $\gamma_\alpha$ as 
\begin{equation}
	\label{eq:mu-c-eq}
	\frac{\Gamma\left(\frac{1}{2}-\frac{1}{\alpha}\right)\Gamma\left(1+\frac{1}{\alpha}
	\right)}{\pi^{3/2}}(2\gamma_\alpha)^{\frac{1}{\alpha}-\frac{1}{2}}+(2\gamma_\alpha)^{-\frac{1}{2}}
	+4\gamma_\alpha^{-\frac{1}{2}+\frac{1}{\alpha}}r'+4\sum_{k=0}^\infty\gamma_\alpha^k r_k=0
\end{equation}
We observe that since $r_k$ are suppressed exponentially upon increasing $k$, we can 
truncate~\eqref{eq:mu-c-eq} by keeping the first $k_\mathrm{max}$ terms in the 
sum. A numerical solution of~\eqref{eq:mu-c-eq} as a function of $\alpha$ is shown 
in Fig.~\ref{fig:gamma-theo}.

\section{Finite-size scaling of the susceptibility $\chi_A^x$}
\label{app:flatX}

Here we derive the flat vector expectation values of the position
correlation matrix $\mathbb{X}_{nm}$ (cf.~\eqref{eq:xx}) given as 
\begin{equation}
	\label{eq:x-corr}
 \mathbb{X}_{nm} = \frac{\sqrt{g}}{2L}\sum_{k=0}^{L-1}
 \frac{e^{\II(n-m)2\pi k/L}}{\sqrt{2\mu + {\omega}_k}} ,
 \quad \text{with} \quad
 \omega_k = [2(1-\cos(2\pi k/L))]^{\frac{\alpha}{2}}.
\end{equation}
We use Poisson's summation formula~\eqref{eq:poisson-1} to decompose the position
correlator into a thermodynamic and a finite-size component, viz.,
\begin{equation}
	\label{eq:X-deco}
	\mathbb{X}_{nm} =\mathbb{X}^{\rm (th)}_{nm} + \mathbb{X}^{\rm (L)}_{nm}.
\end{equation}
Specifically, we have
\begin{align}
\label{eq:Xth-def}
 &\mathbb{X}^{\rm (th)}_{nm} =
 \frac{\sqrt{g}}{2} \int_0^{2\pi} \frac{\D k}{2\pi} \frac{e^{\II (n-m) k}}{\sqrt{2\mu + \omega_k}}\\
 \label{eq:Xfs-def}
 &\mathbb{X}^{\rm (L)}_{nm} =
 \sqrt{g}  \sum_{j=1}^{\infty}\int_0^{2\pi} \frac{\D k}{2\pi} e^{\II (n-m)k}
 \frac{\cos\left( L j k\right)}{\sqrt{2\mu + \omega_k}}.
\end{align}
We consider a bipartition of the chain into two parts as $A\cup B$,
with $B$ the complement of $A$. We denote the size of $A$ as $L_A$ and proceed
to compute the flat-vector expectation value of the position correlation matrix
\begin{equation}
	\label{eq:X-susc}
	\chi^{x}_A=\langle 1|\mathbb{X}|1\rangle_A:=\frac{1}{L_A}\sum_{n,m=0}^{L_A-1}\mathbb{X}_{nm}.
\end{equation}
Notice that $\chi^{x}_A$ has the form of the susceptibility associated to $\mathbb{X}$ 
restricted to subsystem $A$. 
In the following we consider $L_A=L/2$ and treat the thermodynamic and
the finite-size contributions separately. 

\subsection{Thermodynamic contribution}

We observe that Eq.~(\ref{eq:Xth-def}) only depends on the difference $n-m$. Thus we 
can exploit translation invariance using the trivial identity
\begin{equation}
\label{eq:double-sum}
\sum_{n,m=0}^{L/2-1}f(n-m)=\frac{L}{2}\sum_{n=-L/2}^{L/2}\left(1-\frac{2|n|}{L}\right)f(n). 
\end{equation}
We find for the thermodynamic contribution (cf.~\eqref{eq:Xth-def})
\begin{equation}
	\label{eq:step}
	\bra{1}\mathbb{X}^{\rm (th)}\ket{1}_{A}
	=\frac{\sqrt{g}}{2} \int_{0}^{2\pi} \frac{\D k}{2\pi}
	\frac{1}{\sqrt{2\mu + \omega_k}}+\sqrt{g}\sum_{n=1}^{L/2}
\int_{0}^{2\pi} \frac{\D k}{2\pi}
	\frac{\cos(k n)}{\sqrt{2\mu + \omega_k}}
	\left(1-\frac{2n}{L}\right).
\end{equation}
The first term in~\eqref{eq:step} is subleading for large $L$ and is omitted in the following. 
The second term consists of two contributions, which up to a global $\sqrt{g}$
factor read as 
\begin{align}
	T_1 &:= \sum_{n=1}^{L/2}
\int_{0}^{2\pi} \frac{\D k}{2\pi}
	\frac{\cos(k n)}{\sqrt{2\mu + \omega_k}},\\
	T_2 &:=- \frac{2}{L}\sum_{n=1}^{L/2}
\int_{0}^{2\pi} \frac{\D k}{2\pi}
	\frac{n\cos(k n)}{\sqrt{2\mu + \omega_k}}.
\end{align}
We consider the contributions $T_1$ and $T_2$ separately, and proceed as
for the spherical parameter in~\ref{app:mu}. We obtain
\begin{equation}
	\label{eq:split}
T_1 \simeq 2\sum_{n=1}^{L/2}\int_0^{\Lambda}\frac{\D k}{2\pi}\frac{\cos(kn)}{\sqrt{2\mu+k^\alpha}}
	=\sum_{n=1}^{L/2}\int_{c-i\infty}^{c+i\infty} \frac{\D s}{2\pi i}
	\int_{0}^\Lambda\frac{\D k}{\pi}\frac{k^{-s}}{\sqrt{2\mu+k^\alpha}}\cos\left(\frac{\pi}{2}s\right)\Gamma(s)n^{-s}.
\end{equation}
Here we expanded the dispersion  $\omega_k$ around $k=0$. Since the scaling of the entanglement gap is 
determined by the lower part of the dispersion, this approximation will not affect our results. 
The factor two in the first row in~\eqref{eq:split} accounts for the fact that the dispersion $\omega_k$ is 
singular at $k=0$ and $k=2\pi$. The two singularities give the same contributions. Moreover, 
in~\eqref{eq:split} we replaced the integration domain $[0,2\pi]$ with $[0,\Lambda]$, where $\Lambda$ is a cutoff.
Again, as the scaling of the entanglement gap is determined by
the low-energy part of the spectrum of the model, we can neglect contributions that
depend on $\Lambda$.
After performing the sum over $n$ and the integration over $k$ in~\eqref{eq:split}, we obtain
\begin{equation}
	\label{eq:last}
	T_1\simeq
	\frac{\pi^{-\frac{3}{2}}}{\alpha}\int_{c-i\infty}^{c+i\infty} \frac{\D s}{2\pi i}\cos\left(\frac{\pi}{2}s\right)\Gamma(s)
	\Gamma\left(\frac{1-s}{\alpha}\right)\Gamma\left(\frac{1}{2}+\frac{s-1}{\alpha}\right)
	(2\mu)^{-\frac{1}{2}+\frac{1-s}{\alpha}} H_{L/2}(s),
\end{equation}
where we neglect terms that depend on the cutoff $\Lambda$ and consider the limit $\mu\to0$. Here 
$H_{x}(s)$ is the harmonic number~\cite{NIST:DLMF}. 
The inverse Mellin transform is performed by employing the same contour as in Fig.~\ref{fig:contour}. To perform the integral in~\eqref{eq:last}, let us first analyze the singularity 
structure of the integrand. Now, we observe that 
\begin{itemize}
\item $\cos(\pi s/2)\Gamma(s)$ has poles at $s=-2p$ with $p\in\mathbb{N}$, 
		all of which contribute to the integral. Let us define these contributions 
		as $C_{2p}$. 
 \item $\Gamma((1-s)/\alpha)$ has poles for $s\ge 1$ which do not contribute to the integral. 
 \item $\Gamma(1/2+(s-1)/\alpha)$ has poles at $s=1-(2p+1)/2\alpha$, with $p\in\mathbb{N}$ 
	 which do contribute. Let us define these contributions as $C_{2p+1}$. 
 \item The harmonic number $H_{L/2}(s)$ is holomorphic, although in the limit $L\to\infty$
develops a pole at $s=1$. Here we first perform the integration 
in~\eqref{eq:last}, then taking the limit $L\to\infty$.
\end{itemize}
Let us now consider the contributions of the poles.
It is straightforward to check that the contribution 
$C_{2p}$ is given as 
\begin{equation}
	\label{eq:last-1}
	C_{2p}=\frac{\pi^{-\frac{3}{2}}}{\alpha}
	\frac{(-1)^p}{(2p)!}\Gamma\left(\frac{1+2p}{\alpha}\right)
	\Gamma\left(\frac{1}{2}-\frac{2p+1}{\alpha}\right)
	(2\mu)^{-\frac{1}{2}+\frac{1+2p}{\alpha}}
	H_{L/2}(-2p).
\end{equation}
After expanding $H_{L/2}(x)$ for $L\to\infty$ in~\eqref{eq:last-1}, we obtain that 
\begin{equation}
	\label{eq:last-2}
	C_{2p}=\frac{\pi^{-\frac{3}{2}}}{\alpha}
	\frac{(-1)^p}{(2p)!}\Gamma\left(\frac{1+2p}{\alpha}\right)
	\Gamma\left(\frac{1}{2}-\frac{2p+1}{\alpha}\right)
	(2\mu)^{-\frac{1}{2}+\frac{1+2p}{\alpha}}
	\frac{1}{1+2p}
	\left(\frac{L}{2} \right)^{1+2p}.
\end{equation}
In the ferromagnetic phase the spherical parameter scales as $\mu\propto 1/L^2$.
Thus, it is clear from~\eqref{eq:last-2} that 
$C_{2p}\simeq L^{2p(\alpha-2)/\alpha +2-2/\alpha}$.
The exponent $2p(\alpha-2)/\alpha+2-2/\alpha$ decreases upon 
increasing $p$, for any $\alpha$. Thus, by considering
the case with $p=0$, we find the leading exponent to be $2-2/\alpha < \alpha/2$.
Conversely, at the critical point, the spherical parameter scales as $\mu\simeq L^{-\alpha}$.
It is straightforward to check that this scaling implies that~\eqref{eq:last-2} scales as
$\simeq L^{\frac{\alpha}{2}}$ for any $p$.

Let us now consider the contribution $C_{2p+1}$. 
From Eq.~\eqref{eq:last} this  reads
\begin{equation}
	\label{eq:s3}	
	C_{2p+1}=\frac{(-1)^p}{p! \pi^{\frac{3}{2}}}
	\sin\left[\frac{\pi}{2}\left(p+\frac{1}{2}\right)\alpha\right]
	\Gamma\left[1-\left(p+\frac{1}{2}\right)\alpha\right]
	\Gamma\left(p+\frac{1}{2}\right)(2\mu)^{p}
	H_{L/2}\left[1-\left(p+\frac{1}{2}\right)\alpha\right].
\end{equation}
Again, after expanding $H_{L/2}(x)$ for large $L$, we find
\begin{equation}
	\label{eq:s4}
	C_{2p+1}=\frac{2(-1)^p}{\alpha p! 
	\pi^{\frac{3}{2}}(2p+1)}\sin\left[\frac{\pi}{2}\left(p+\frac{1}{2}\right)\alpha\right]
	\Gamma\left[1-\left(p+\frac{1}{2}\right)\alpha\right]
	\Gamma\left(p+\frac{1}{2}\right)(2\mu)^{p}
	\left( \frac{L}{2} \right)^{\left(p+\frac{1}{2}\right)\alpha}.
\end{equation}
In the ferromagnetic region Eq.~\eqref{eq:s4} gives  
$C_{2p+1}\simeq L^{\frac{\alpha}{2}+(\alpha-2)p}$. Again, the leading 
behavior is obtained for $p=0$. Moreover, 
at criticality one has $\propto L^\frac{\alpha}{2}$. Overall we find
\begin{multline}
	\label{eq:T1-res}
	T_1 \simeq \sum_{p=0}^\infty
	\frac{\pi^{-\frac{3}{2}}}{\alpha}\frac{(-1)^p}{2p+1}
	\Bigg\{
	\frac{1}{(2p)!}\Gamma\left(\frac{1+2p}{\alpha}\right)
	\Gamma\left(\frac{1}{2}-\frac{2p+1}{\alpha}\right)
	(2\mu)^{-\frac{1}{2}+\frac{1+2p}{\alpha}}
	\left(\frac{L}{2} \right)^{1+2p}\\
	+ \frac{2}{p!}\sin\left[\frac{\pi}{2}
	\left(p+\frac{1}{2}\right)\alpha\right]
	\Gamma\left[1-\left(p+\frac{1}{2}\right)\alpha\right]
	\Gamma\left(p+\frac{1}{2}\right)(2\mu)^{p}
\left( \frac{L}{2} \right)^{\left(p+\frac{1}{2}\right)\alpha}\Bigg\}.
\end{multline}
The leading part can be retrieved for $p=0$, viz.,
\begin{equation}
T_1\simeq \frac{\pi^{-\frac{3}{2}}}{\alpha}
	\Bigg[\Gamma\left(\frac{1}{\alpha}\right)
	\Gamma\left(\frac{1}{2}-\frac{1}{\alpha}\right)
	(2\mu)^{-\frac{1}{2}+\frac{1}{\alpha}}
	\frac{L}{2}
	+ 2\sqrt{\pi}\sin\left(\frac{\pi}{4}\alpha\right)
\Gamma\left(1-\frac{\alpha}{2}\right)
	\left( \frac{L}{2} \right)^{\frac{\alpha}{2}}\Bigg].
\end{equation}
Let us now discuss the second term in~\eqref{eq:step}. This is 
treated in the same way as the first one. The only difference is that 
in doing  the Mellin inverse transform, one has to shift  by one to the left 
the contour in  Fig.~\ref{fig:contour}. This is due to the multiplying $n$ factor in
the sum in~\eqref{eq:step}. Hence, we find  
\begin{equation}
	\label{eq:sec}
	T_2 \simeq \frac{2\pi^{-\frac{3}{2}}}{\alpha L}\int_{c-i\infty}^{c+i\infty}
	\frac{\D s}{2\pi i} \sin\left(\frac{\pi}{2}s\right)\Gamma(s+1)
	\Gamma\left(\frac{1}{2}+
	\frac{s}{\alpha}\right)\Gamma\left(-\frac{s}{\alpha}\right)
	(2\mu)^{-\frac{1}{2}-\frac{s}{\alpha}}H_{L/2}(s),
\end{equation}
with $-\frac{\alpha}{2}<c<0$.
Similar to the treatment of the  term $T_1$, we identify the relevant poles
to compute the contour integral at $s= -(2p+1)$ and  
$s =-(2p+1)\alpha/2 $.
Let us define as $C_{2p+1}'$ the  
contribution to Eq.~\eqref{eq:sec} from the poles at $s=-(2p+1)$. This  
reads
\begin{equation}
	C_{2p+1}'\simeq\frac{2\pi^{-\frac{3}{2}}}{\alpha L}
	\frac{(-1)^{p+1}}{(2p)!} \Gamma\left(\frac{1}{2}-
	\frac{2p+1}{\alpha}\right)
	\Gamma\left(\frac{2p+1}{\alpha}\right)
	(2\mu)^{-\frac{1}{2}+(2p+1)/\alpha} H_{L/2}(-(2p+1)).
\end{equation}
In the large $L$ limit the leading scaling of this contribution is
\begin{equation}
	C'_{2p+1}\simeq\frac{\pi^{-\frac{3}{2}}}{2\alpha}
	\frac{(-1)^{p+1}}{ (2p)!}\Gamma\left(\frac{1}{2}
	-\frac{2p+1}{\alpha}\right)
	\Gamma\left(\frac{2p+1}{\alpha}\right)
	(2\mu)^{-\frac{1}{2}+(2p+1)/\alpha}\left(\frac{L}{2}\right)^{2p+1}
	\frac{1}{1+p}.
\end{equation}
In the ordered phase  one has that  $ C'_{2p+1}\simeq L^{1+(\alpha-2)(2p+1)/\alpha}$.
We again notice that the exponent is always smaller than $\alpha/2$, 
and it decreases with increasing $p$, meaning that larger $p$ corresponds to 
smaller contributions.
At criticality we find that  $C'_{2p+1}\simeq L^\frac{\alpha}{2}$, irrespective of 
$p$. 

Let us now consider the contribution $C_{2p+1}''$ of the 
poles at $s=-(2p+1)\alpha/2$. Their contribution to the integral in Eq.~(\ref{eq:sec}) is
\begin{equation}
	C_{2p+1}''\simeq 2\frac{(-1)^{p+1}}{p!L\pi^{\frac{3}{2}}}
	\sin\left[\frac{\pi\alpha}{2}\left(p+\frac{1}{2}\right)\right]
	\Gamma\left[1-\left(p+\frac{1}{2}\right)\alpha\right]
	\Gamma\left(p+\frac{1}{2}\right)(2\mu)^{p}
	H_{L/2}\left[-\left(p+\frac{1}{2}\right)\alpha\right].
\end{equation}
Again, after expanding the harmonic number $H_{L/2}(s)$ in the large $L$ limit, we have
\begin{equation}
	C_{2p+1}''\simeq 2\frac{(-1)^{p+1}}
	{p!\pi^{\frac{3}{2}}}\sin\left[\frac{\pi}{2}\left(p+\frac{1}{2}\right)\alpha\right]
	\Gamma\left[1-\left(p+\frac{1}{2}\right)\alpha\right]
	\Gamma\left(p+\frac{1}{2}\right)(2\mu)^{p}
	\frac{\left(L/2\right)^{(p+1/2)\alpha}}{2+(2p+1)\alpha}.
\end{equation}
In the ferromagnetic phase one has that 
$C_{2p+1}''\simeq L^{\frac{\alpha}{2}+p(\alpha-2)}$, whereas at 
criticality one has  $C_{2p+1}''\simeq L^\frac{\alpha}{2}$.
Putting everything together, we obtain 
\begin{multline}
	\label{eq:T2-res}
	T_2 \simeq
 \sum_{p=0}^\infty \frac{(-1)^{p+1}}{\pi^{\frac{3}{2}}}\Bigg\{
 	\frac{1}{2\alpha}
	\frac{1}{ (2p)!}\Gamma\left(\frac{1}{2}
	-\frac{2p+1}{\alpha}\right)
	\Gamma\left(\frac{2p+1}{\alpha}\right)
	(2\mu)^{-\frac{1}{2}+(2p+1)/\alpha}
	\frac{\left(L/2\right)^{2p+1}}{1+p} 
 \\+
 \frac{2}{p!}\sin\left[\frac{\pi}{2}\left(p+\frac{1}{2}\right)\alpha\right]
 \Gamma\left[1-\left(p+\frac{1}{2}\right)\alpha\right]
	\Gamma\left(p+\frac{1}{2}\right)(2\mu)^{p}
\frac{\left(L/2\right)^{(p+1/2)\alpha}}{2+(2p+1)\alpha}\Bigg\}. 
\end{multline}
Finally, we should stress that in deriving  $T_1$ and $T_2$ we considered the 
limit $\mu\to0$. This allowed us to neglect all the cutoff-dependent contributions. 
At the critical point all the contributions~\eqref{eq:T1-res} and~\eqref{eq:T2-res} 
are of the same order $L^{\alpha/2}$ in the large $L$ limit. 
They encode universal information about 
the critical behavior of the system. On the other hand, 
in the ordered phase, the large-$L$ behavior of the different terms in~\eqref{eq:T1-res} 
and~\eqref{eq:T2-res} depends on $p$. Specifically, larger $p$ corresponds to more 
suppressed contributions. As a consequence,  in the ferromagnetic phase some of the 
terms in~\eqref{eq:T1-res} and~\eqref{eq:T2-res} for large enough $p$ can be 
subleading as compared with the cutoff-dependent terms that we neglected. However, 
it is crucial to stress that the leading behavior of $T_1$ and $T_2$ is determined 
by the terms with $p=0$ in~\eqref{eq:T1-res} and~\eqref{eq:T2-res}.

\subsection{Finite-size contribution}
\label{sec:fs-app}

Let us consider the finite-size contribution to $\langle1|\mathbb{X}|1\rangle_A$, 
which corresponds to the second term in the decomposition in~\eqref{eq:X-deco}. 
We recall that it is given as (cf.~\eqref{eq:Xfs-def}) 
\begin{equation}
	\bra{1}\mathbb{X}^{\mathrm{(L)}}\ket{1}_{A} =
	\frac{2\sqrt{g}}{L}\sum_{j=1}^\infty\sum_{n,m=0}^{L/2}\int_0^{2\pi}\frac{\D k}{2\pi}e^{ik(n-m)}\frac{\cos(Ljk)}{\sqrt{2\mu+\omega_k}}.
\end{equation}
This can be rewritten as 
\begin{equation}
	\label{eq:me-x}
	\bra{1}\mathbb{X}^{\mathrm{(L)}}\ket{1}_{A} \simeq
	\frac{2\sqrt{g}}{L}\sum_{j=1}^\infty\sum_{n,m=0}^{L/2}\int_{0}^{\Lambda}
	\frac{\D k}{2\pi}\left(e^{ik(n-m+Lj)}+e^{ik(n-m-Lj)}\right)\frac{1}{\sqrt{2\mu+k^\alpha}},
\end{equation}
where we expanded the dispersion at small $k$, we introduced the cutoff $\Lambda$,  
and we multiplied the result by a factor two to account for the singularity at $k=0,2\pi$. 
To proceed, we use that the  Mellin transform of $e^{ikx}$ with respect to $x$ 
is $(-ik)^{-s}\Gamma(s)$. Thus, we can rewrite~\eqref{eq:me-x} to obtain
\begin{multline}
	\label{eq:fs-step}
	\bra{1}\mathbb{X}^{\mathrm{(L)}}\ket{1}_{A} \simeq
	\frac{\sqrt{g}}{L\pi^{\frac{3}{2}}\alpha}\sum_{j=1}^\infty\sum_{n,m=0}^{L/2}
	\int_{c-i\infty}^{c+i\infty}\frac{\D s}{2\pi i}
	(2\mu)^{-\frac{1}{2}+\frac{1-s}{\alpha}}\Gamma(s)\Gamma\left(\frac{1-s}{\alpha}\right)\\
	\times \, \Gamma\left(\frac{1}{2}
	+\frac{s-1}{\alpha}\right)\frac{(-i)^{-s}}{(n-m\pm jL)^{s}},
\end{multline}
where we sum over the $\pm$ in the last term, and we choose $1-\alpha/2<c<1$.
Now, we carry out the sum over $j$.  This step, however, requires $c>1$. After noticing that 
the pole at $s=1$ in~\eqref{eq:fs-step} is removed by the double sum, we 
can shift the contour across the pole to the right without additional contributions. 
Using Eq.~\eqref{eq:double-sum} and dropping
the subleading contribution for $p=0$ allows us to rewrite~\eqref{eq:fs-step} as 
\begin{multline}
	\label{eq:fs-step-2}
	\bra{1}\mathbb{X}^{\mathrm{(L)}}\ket{1}_{A} \simeq
	\frac{\sqrt{g}}{\pi^{\frac{3}{2}} \alpha}\sum_{j=1}^\infty\sum_{r=1}^{L/2}
	\int_{c-i\infty}^{c+i\infty}\frac{\D s}{2\pi i}
	(2\mu)^{-\frac{1}{2}+\frac{1-s}{\alpha}}\Gamma(s)\Gamma\left(\frac{1-s}{\alpha}\right)\\
	\times \, \Gamma\left(\frac{1}{2}	+\frac{s-1}{\alpha}\right)
	\left[1-2\frac{r}{L}\right]\frac{\cos(\pi s/2)}{(r\pm jL)^{s}}.
\end{multline}
Again, the integrand is regular at $s=1$ and we moved the integration contour considering 
$1<c<1+\alpha$. After carrying out  the infinite $j$ sum, we find 
\begin{multline}
\label{eq:fs-mellin}
\bra{1}\mathbb{X}^{\mathrm{(L)}}\ket{1}_{A}\simeq
	\sum_{r=1}^{L/2}
	\frac{\sqrt{g}}{\pi^{\frac{3}{2}}\alpha}\int_{c-i\infty}^{c+i\infty}\frac{\D s}{2\pi i}
	(2\mu)^{-\frac{1}{2}+\frac{1-s}{\alpha}}L^{-s}
	\Gamma(s)\Gamma\left(\frac{1-s}{\alpha}\right)\Gamma\left(\frac{1}{2}+\frac{s-1}{\alpha}
	\right)\\
	\times \, \left[1-2\frac{r}{L}\right]\cos\left(\frac{\pi}{2} s\right)\zeta\left(s,1\pm\frac{r}{L}\right),
\end{multline}
where $\zeta(s,a)$ is the Hurwitz zeta function~\cite{NIST:DLMF}.
The structure of the poles in~\eqref{eq:fs-mellin} is similar to that 
found for the  spherical parameter (see~\ref{app:mu} ). 
For the following it is important to stress that  
the Hurwitz zeta functions have a simple pole at $s=1$ with residue one. 
The pole of $\zeta(s,a)$ gives  the leading contribution of the integral~\eqref{eq:fs-mellin} 
at $L\to\infty$. 
Specifically, we have 
\begin{equation}
	\label{eq:mu-nm}
	C_H=\frac{1}{4}\sqrt{\frac{g}{2\mu}}\left(1-\frac{2}{L}\right). 
\end{equation}
Here  we can neglect the  $1/L$ term because it is subleading at large $L$. 
Eq.~\eqref{eq:mu-nm} at criticality is ${\mathcal O}(L^{-\alpha/2})$, whereas in the 
ordered phase it is ${\mathcal O}(L^{-1})$. 

Let us now denote as $C_{2p+1}'''$ the contributions of the  poles 
at $s=1-(2p+1)\alpha/2$. One obtains
\begin{multline}
	\label{eq:first-corr}
	C'''_{2p+1}\simeq\frac{\sqrt{g}}{\pi^\frac{3}{2}}\sum_{r=1}^{L/2}
	\frac{(-1)^p}{p!}(2\mu)^{p}
	L^{-1+\frac{2p+1}{2}\alpha}\left[1-2\frac{r}{L}\right]
	\sin\left(\frac{\pi}{2}\alpha(p+1/2)\right)\\
	\times\, \Gamma(1-(p+1/2)\alpha)
	\Gamma\left(p+1/2\right)\zeta\left(1-\frac{2p+1}{2}\alpha,1\pm\frac{r}{L}\right),
\end{multline}
At criticality we have $C_{2p+1}'''={\mathcal O}(L^{\alpha/2})$ for any $p$, whereas 
in the ordered phase terms with larger $p$ are more suppressed in the large $L$ limit. 
If we are interested only in the leading term in~\eqref{eq:first-corr}, i.e., 
for $p=0$, we can replace the sum over $r$ in~\eqref{eq:first-corr} with an integral, to obtain
\begin{equation}
	\label{eq:l-sub}
	C'''_{1}\simeq 
	\frac{\sqrt{g}L^{\frac{\alpha}{2}}}{2\pi^\frac{3}{2}}\sin\left(\frac{\pi\alpha}{4}\right)
	\Gamma\left(1-\frac{\alpha}{2}\right)\Gamma\left(\frac{1}{2}\right)
	\int_0^1 \D x
	(1-x)\zeta\left(1-\frac{\alpha}{2},1\pm \frac{x}{2}\right). 
\end{equation}
Let us now consider the contribution $C_{2p}'''$ of the poles at $s=-2p$. 
We remark that  these poles do not contribute to 
the finite-size scaling of the spherical parameter 
(see section~\ref{app:mu}) because $\zeta(-2p)=0$ for any $p$, i.e., the residue is 
zero. However, here they give 
a nonzero contribution. One obtains 
\begin{equation}
	\label{eq:sec1}
	C_{2p}^{'''}=\frac{1}{2\pi^\frac{3}{2}\alpha}
	\sum_{r=1}^{L/2} \frac{(-1)^n}{(2p)!}
	(2\mu)^{-\frac{1}{2}+\frac{2p+1}{\alpha}}
	L^{2p-1}(L-2r)
	\Gamma\left(\frac{1-s_p}{\alpha}\right)
	\Gamma\left(\frac{1}{2}+\frac{s_p-1}{\alpha}\right)
	\zeta\left(-2p,1\pm\frac{r}{L}\right). 
\end{equation}
Again, the contribution $C'''_{2p}$ decreases upon increasing $p$. 
The leading term corresponds to $p=0$. This, however, is subleading 
compared to~\eqref{eq:l-sub} in the ordered phase. At criticality the contribution~\eqref{eq:sec1} 
is ${\mathcal O}(L^{\alpha/2})$ for any $p$. 
\begin{figure}[t]
 \centering
 \includegraphics[width = .495\textwidth]{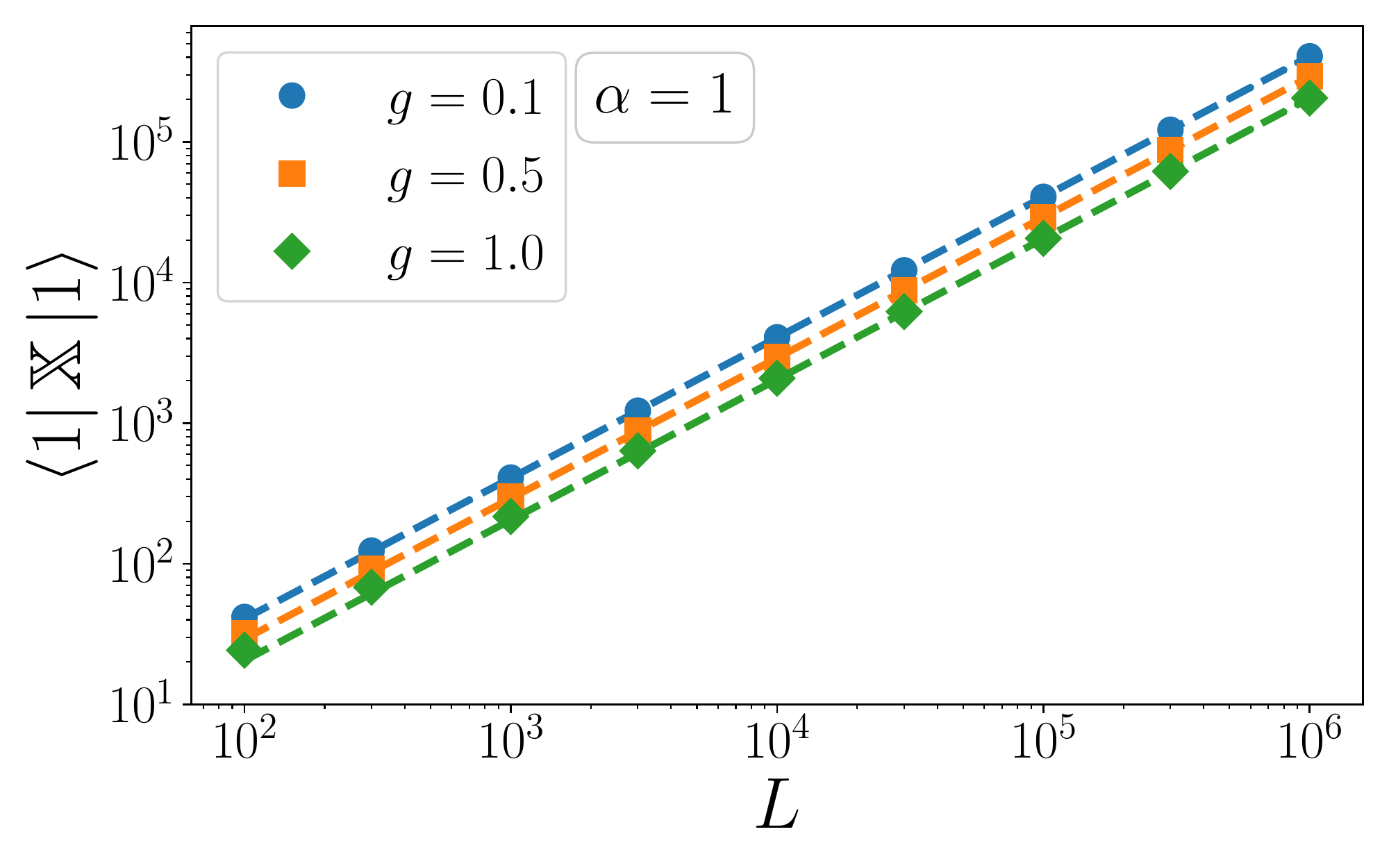}
 \includegraphics[width = .495\textwidth]{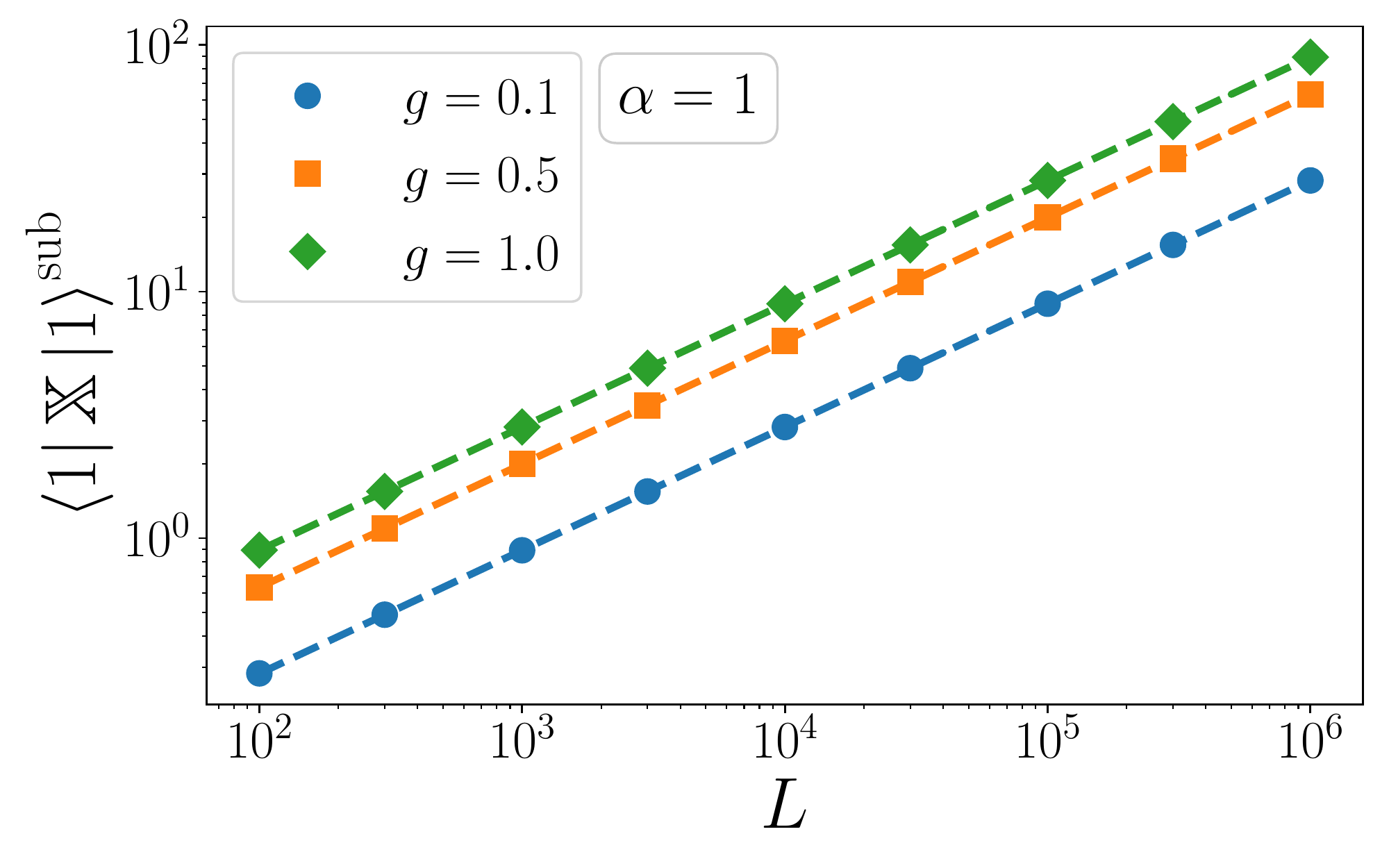}
 \includegraphics[width = .495\textwidth]{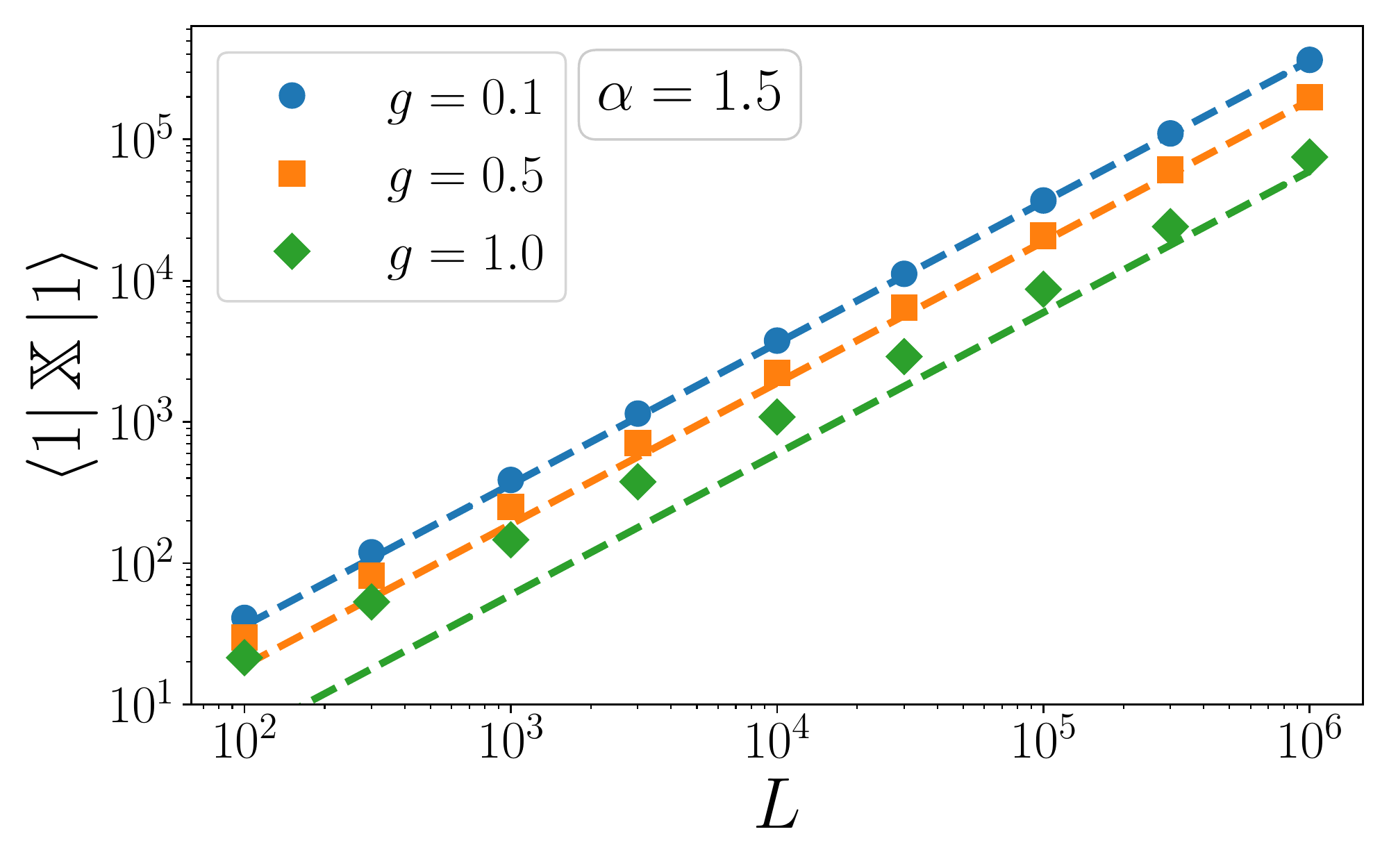}
 \includegraphics[width = .495\textwidth]{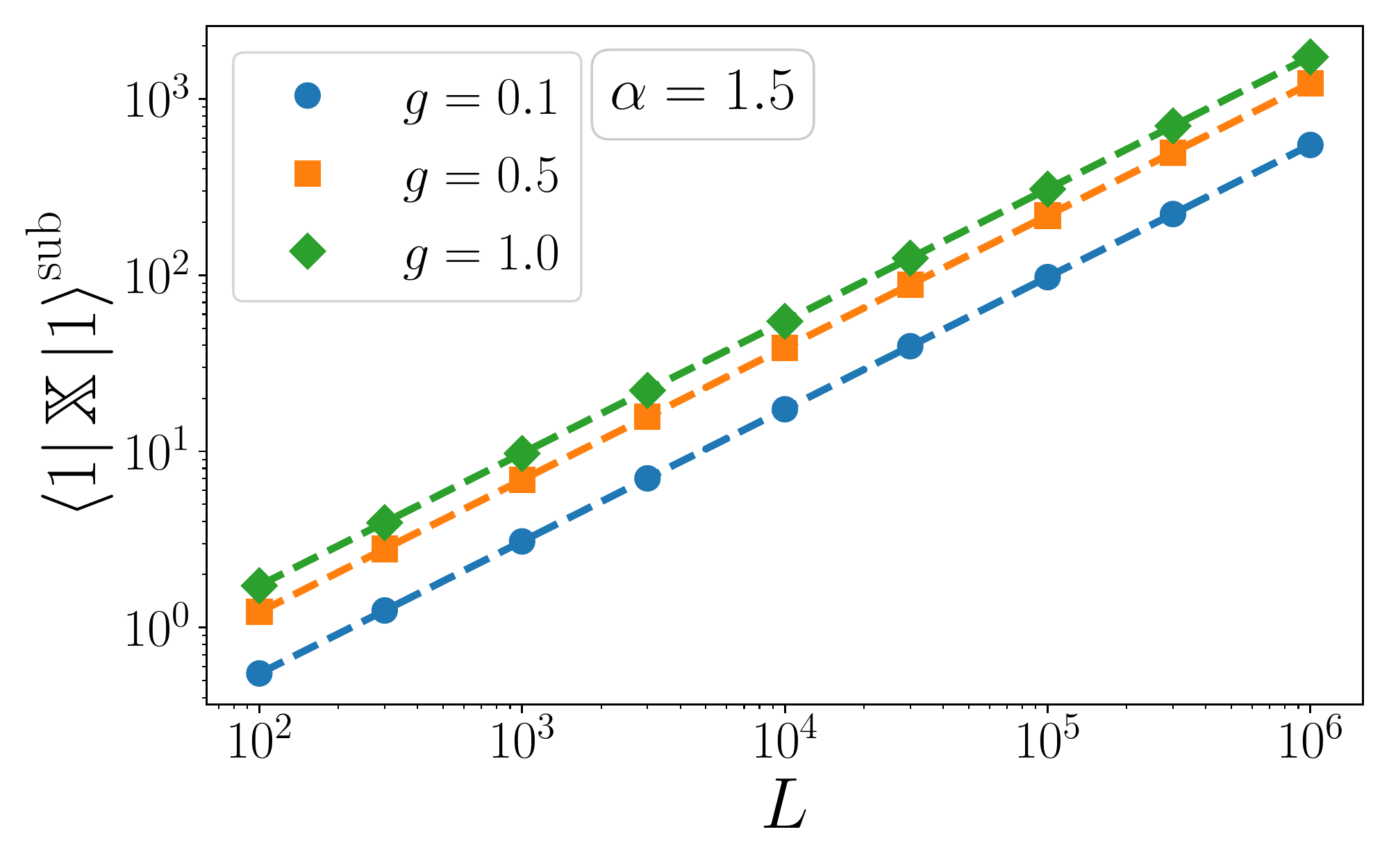}
 \caption{ Finite-size scaling of $\chi_A^x:=\langle 1|\mathbb{X} 
 |1\rangle$ in the ferromagnetic phase of the quantum spherical model with long-range 
 interactions. In the figure we plot $\langle1|\mathbb{X}|1\rangle_A$ 
 versus $L$. Notice the logarithmic scale on both axes. 
 (Top row).  Results for $\alpha=1$. 
 In the left panel we focus on the the leading scaling behavior in the large $L$ limit. 
 The different symbols correspond to different values of $g$. The lines are the 
 analytic results (first term in~\eqref{eq:chix-an}). The right panel shows the 
 first subleading term $\langle1|\mathbb{X}|1\rangle_A^{\mathrm{sub}}$ of 
 $\left<1\right|\mathbb{X}\left|1\right>$. The data are obtained from those in 
 the left panel by subtracting the analytic prediction for the leading behavior. 
 The dashed line are the analytic results (second term in~\eqref{eq:chix-an}). 
 (Bottom row). The same as in the top row for $\alpha=1.5$. 
 }
 \label{fig:flatX}
\end{figure}
%
%

\section{Finite-size scaling of the susceptibility $\chi_A^t$}
\label{app:pcorr}

\noindent
Here we derive the flat-vector expectation values of the momentum
correlation matrix $\langle1|\mathbb{P}|1\rangle_A$, i.e., of the 
susceptibility $\chi_A^t$. The correlation matrix $\mathbb{P}_{nm}$ reads (see Eq.~\eqref{eq:pp}) 
\begin{equation}
	\label{eq:p-app}
 \mathbb{P}_{nm} = \frac{1}{\sqrt{g}}\frac{1}{2L}\sum_{k=0}^{L-1}
 e^{\II(n-m)\frac{2\pi}{L}k}\sqrt{2\mu + \omega_k}, 
\end{equation}
with the frequency $\omega_k$ defined as in~\eqref{eq:qsm-ham}. 
Again, we use Poisson's summation formula~\eqref{eq:poisson-1} to 
split~\eqref{eq:p-app} into a thermodynamic and a finite-size part, i.e.,
$\mathbb{P}_{nm} =\mathbb{P}^{\rm (th)}_{nm} + \mathbb{P}^{\rm (L)}_{nm}$.
Specifically, we have
\begin{align}
\label{eq:Pth-def}
 \mathbb{P}^{\rm (th)}_{nm} &=
 \frac{1}{2}\frac{1}{\sqrt{g}} \int_0^{2\pi} \frac{\D k}{2\pi} e^{\II (n-m) k}\sqrt{2\mu + \omega_k},\\
 \label{eq:Pfs-def}
 \mathbb{P}^{\rm (L)}_{nm} &=
 \frac{1}{\sqrt{g}}  \sum_{j=1}^{\infty}\int_0^{2\pi} \frac{\D k}{2\pi} e^{\II (n-m)k}
\cos\left( L j k\right)\sqrt{2\mu + \omega_k}.
\end{align}
We consider a bipartition of the chain into two parts as $A\cup B$,
with $B$ the complement of $A$. We denote the size of $A$ as $L_A$ and proceed
to compute the flat-vector expectation value of the momentum correlation matrix
\begin{equation}
	\chi_A^t=\langle 1|\mathbb{P}|1\rangle_A:=\frac{1}{L_A}\sum_{n,m=0}^{L_A-1}\mathbb{P}_{nm}.
\end{equation}
In the following we consider $L_A=L/2$ and treat the thermodynamic and
the finite-size contributions separately.

%

\subsection{A useful integral}
\label{app:useful}
In order to extract the finite-size scaling  
of $\bra{1}\mathbb{P}\ket{1}_A$ we need to analyze
the ``universal'' part of the integral
\begin{equation}
	\label{eq:u-int}
 \mathfrak{J}(s) = \int_0^{2\pi} \frac{\D k}{2\pi}k^{-s} \sqrt{2\mu + \omega(k)}.
\end{equation}
Hence, it suffices to consider the small $k$ limit and study $\mu\to0$. 
To this end, we introduce a cutoff $\Lambda$ as follows
\begin{equation}
 \mathfrak{J}(s) \simeq \int_0^\Lambda \frac{\D k}{\pi} k^{-s} \sqrt{2\mu + \omega(k)}=
 \frac{1}{\pi}\frac{\Lambda^{1-s}}{1-s} \sqrt{2\mu}+
 \int_0^\Lambda \frac{\D k}{\pi} k^{-s} \left(\sqrt{2\mu + \omega(k)} - \sqrt{2\mu}\right). 
\end{equation}
After using the short wavelength approximation and after 
changing variable as $y^2 = k^\alpha/(2\mu)$, we obtain 
\begin{multline}
 \mathfrak{J}(s) \simeq
 \frac{1}{\pi}\frac{\Lambda^{1-s}}{1-s} \sqrt{2\mu}+
 \frac{2}{\alpha}\int_0^{\sqrt{\Lambda^\alpha/2\mu}} 
\frac{\D y}{\pi} y^{\frac{2}{\alpha}(1-s)-1} \left(\sqrt{1 + y^2} -1\right)
(2\mu)^{\frac{1}{2}+\frac{1-s}{\alpha}}
\\
\simeq
\frac{2}{\alpha}(2\mu)^{\frac{1}{2}+\frac{1-s}{\alpha}}\int_0^{\infty} 
\frac{\D y}{\pi} y^{\frac{2}{\alpha}(1-s)-1} \left(\sqrt{1 + y^2} -1\right)
\end{multline}
where we took the limit $\mu\to0$, we neglected all cutoff-dependent 
contributions and we multiplied by two the result to account for the 
singularities.
The remaining integral is readily evaluated, and we find for 
$1+\alpha/2 < \operatorname{Re} (s) < 1+\alpha$
\begin{equation}\label{eq:Pintegral}
	\mathfrak{J} \simeq -\frac{\pi^{-3/2}}{2\alpha} (2\mu)^{\frac{1}{2} + \frac{1-s}{\alpha}}
 \Gamma\left(-\frac{1}{2} - \frac{1-s}{\alpha}\right)\Gamma\left(\frac{1-s}{\alpha}\right).
\end{equation}
Eq.~\eqref{eq:Pintegral} contains full information about the 
universal contributions at criticality. One should observe that 
the leading behavior of 
thermodynamic contribution $\langle1|\mathbb{P}^{\mathrm{(th)}}|1\rangle_A$ in 
the large $L$ limit is not ``universal'', meaning that it depends on the 
cutoff $\Lambda$. Cutoff-independent terms are subleading. 
This is in contrast with $\chi_A^x$ (see~\ref{app:flatX}). 

\subsection{Thermodynamic contribution}
\label{app-p-th}

As in~\ref{app:flatX}, we again observe that Eq.~\eqref{eq:Pth-def} 
only depends on the difference $n-m$ and thus, we can rewrite it using Eq.~\eqref{eq:double-sum} 
as
\begin{equation}
	\label{eq:stepp}
	\bra{1}\mathbb{P}^{\rm (th)}\ket{1}_{A}
	=\frac{1}{\sqrt{g}} 
	\int_{0}^{2\pi} \frac{\D k}{2\pi}
	\sqrt{2\mu + \omega_k}\left[\frac{1}{2}+\sum_{n=1}^{L/2}
	\cos(k n)\left(1-\frac{2n}{L}\right)\right].
\end{equation}
As for $\chi_A^x$, we shall treat the three  contributions in the bracket 
separately. For the first contribution in~\eqref{eq:stepp} we find
\begin{equation}
	\label{eq:p-zero}
	\frac{1}{2}\int_{0}^{2\pi} \frac{\D k}{2\pi}
	\sqrt{2\mu + \omega_k}
	\simeq
	A + B \cdot (2\mu) + C \cdot (2\mu)^{\frac{1}{2} + 1/\alpha}, 
\end{equation}
with 
\begin{align}
	\label{eq:A-def}
A &= \int_{0}^{2\pi} \frac{\D k}{4\pi} \sqrt{\omega_k}
   = 2^{\frac{\alpha}{2}-1} \frac{\Gamma((2+\alpha)/4)}{\sqrt{\pi}\Gamma(1+\alpha/4)}\\
   \label{eq:B-def}
B &= \int_{0}^{2\pi} \frac{\D k}{8\pi} \frac{1}{\sqrt{\omega_k}}
   = 2^{-2-\frac{\alpha}{2}} \frac{\Gamma((2-\alpha)/4)}{\sqrt{\pi}\Gamma(1-\alpha/4)}\\
   \label{eq:C-def}
C &= -\frac{\pi^{-3/2}}{4\alpha} 
 \Gamma\left(-\frac{1}{2} - \frac{1}{\alpha}\right)\Gamma\left(\frac{1}{\alpha}\right)
\end{align}
In deriving~\eqref{eq:p-zero} we expanded the integrand for $\mu\to0$, keeping only terms 
up to ${\mathcal O}(\mu)$. This gives the first two terms in~\eqref{eq:p-zero}. 
As it is clear from~\eqref{eq:A-def} and~\eqref{eq:B-def} the prefactors $A$ and $B$ depend 
on the full dispersion $\omega_k$, and hence on the cutoff $\Lambda$. This means that 
the first tow contributions in~\eqref{eq:p-zero} are not ``universal''. The last term in~\eqref{eq:p-zero} is obtained from~\eqref{eq:Pintegral} by fixing $s=0$. This last term depends only on the 
low-energy part of the dispersion, and hence is ``universal''. 

Let us now evaluate the second contribution $T_1$ in~\eqref{eq:stepp}, i.e., 
\begin{equation}
	\label{eq:t1}
	T_1 =\sum_{r=1}^{L/2} \int_{0}^{2\pi} \frac{\D k}{2\pi}
	\sqrt{2\mu + \omega_k}\cos(k r).
\end{equation}
Here we omit the $1/\sqrt{g}$ as compared with~\eqref{eq:stepp}. 
To evaluate~\eqref{eq:t1} we use the 
Mellin technique as in~\ref{app:flatX}. To this end we use the identity 
\begin{equation}
	\label{eq:cos}
	\cos(x) = \int_\gamma \frac{\D s}{2\pi \II} x^{-s} \cos\left(\frac{\pi}{2}s\right)\Gamma(s), 
\end{equation}
Here $\gamma$ denotes a contour in the complex plane enclosing the entire negative 
real axis, and not exceeding $\mathrm{Re}(s)=1$. Thus, Eq.~\eqref{eq:cos} 
can be verified by using Cauchy's residue theorem. 
After carrying out the sum over $r$ in~\eqref{eq:t1}, and subsequently 
expanding the harmonic numbers for $L\to\infty$ yields
\begin{equation}
\label{eq:t1-2}
T_1 \simeq \int_\gamma \frac{\D s}{2\pi \II}
\left(\frac{L}{2}\right)^{1-s}\frac{\cos\left(\pi s/2\right)}{1-s} \Gamma(s) \mathfrak{J}(s),
\end{equation}
where $\mathfrak{J}(s)$ is the integral in~\eqref{eq:u-int}. 
Since the pole at $s=1$ in~\eqref{eq:t1-2} is removed by the 
vanishing of the cosine, we can deform the path $\gamma$ into a new 
path $\gamma'$ that still encloses 
the entire negative axis but closes such that 
$1+\alpha/2<\operatorname{Re}(s)<1+\alpha$. Now, we can use the 
expression in Eq.~\eqref{eq:Pintegral} to obtain 
\begin{equation}
	\label{eq:t1-3}
T_1 \simeq -\frac{\pi^{-\frac{3}{2}}}{2\alpha}\int_{\gamma'} \frac{\D s}{2\pi \II}
\left(\frac{L}{2}\right)^{1-s}(2\mu)^{\frac{1}{2} + \frac{1-s}{\alpha}}
\frac{\cos\left(\pi s/2\right)}{1-s} \Gamma(s) 
\Gamma\left(-\frac{1}{2} - \frac{1-s}{\alpha}\right)\Gamma\left(\frac{1-s}{\alpha}\right). 
\end{equation}
The leading contribution to $T_1$ is readily found from the residue at $s=1+\alpha/2$, i.e., 
\begin{equation}
	\label{eq:t1-fin-1}
	T_1 \simeq \frac{2^{1+\frac{\alpha}{2}}}{\alpha \pi} \sin\left(\frac{\pi}{4}\alpha\right)
	\Gamma\left(1+\frac{\alpha}{2}\right) L^{-\frac{\alpha}{2}}. 
\end{equation}
Subleading contributions can be found from the remaining residues of the integrand in~\eqref{eq:t1-3}. A similar procedure allows us to evaluate the last contribution in~\eqref{eq:stepp}, i.e.,
\begin{multline}
	T_2 = \frac{2}{L}\sum_{n=1}^{L/2} \int_0^{2\pi} \frac{\D k}{2\pi} n\cos( n k ) 
	\sqrt{2\mu+\omega(k)}\\
    \simeq \frac{\pi^{-\frac{3}{2}}}{2\alpha}\int_{\gamma'} \frac{\D s}{2\pi \II}
\left(\frac{L}{2}\right)^{1-s}(2\mu)^{\frac{1}{2} + \frac{1-s}{\alpha}}
\frac{\cos\left(\pi s/2\right)}{s-2} \Gamma(s) 
 \Gamma\left(-\frac{1}{2} - \frac{1-s}{\alpha}\right)\Gamma\left(\frac{1-s}{\alpha}\right).
\end{multline}
Again, the leading contribution comes from the pole at $s=1+\alpha/2$ and we find
\begin{equation}
	\label{eq:t1-fin-2}
	T_2\simeq 
	-\frac{2^{\frac{\alpha}{2}+1}}{\pi} \frac{\sin\left(\pi\alpha/4\right)}{2-\alpha}
	\Gamma\left(1+\frac{\alpha}{2}\right) L^{-\frac{\alpha}{2}}. 
\end{equation}
Finally, by putting together~\eqref{eq:t1-fin-1} and~\eqref{eq:t1-fin-2} we 
obtain the result for $\langle1|\mathbb{P}^{\mathrm{(th)}}|1\rangle_A$ as
\begin{equation}
	\label{eq:pth-fin}
	\bra{1}\mathbb{P}^{({\rm th})}\ket{1}_A\simeq
	\frac{2^{1+\frac{\alpha}{2}}}{\pi\sqrt{g}}\sin\left(\frac{\pi}{4}\alpha\right)
	\Gamma\left(1+\frac{\alpha}{2}\right) \frac{2}{\alpha(2-\alpha)} L^{-\frac{\alpha}{2}}. 
\end{equation}
%
%

%
%
\begin{figure}[t]
 \centering
 \includegraphics[width = .495\textwidth]{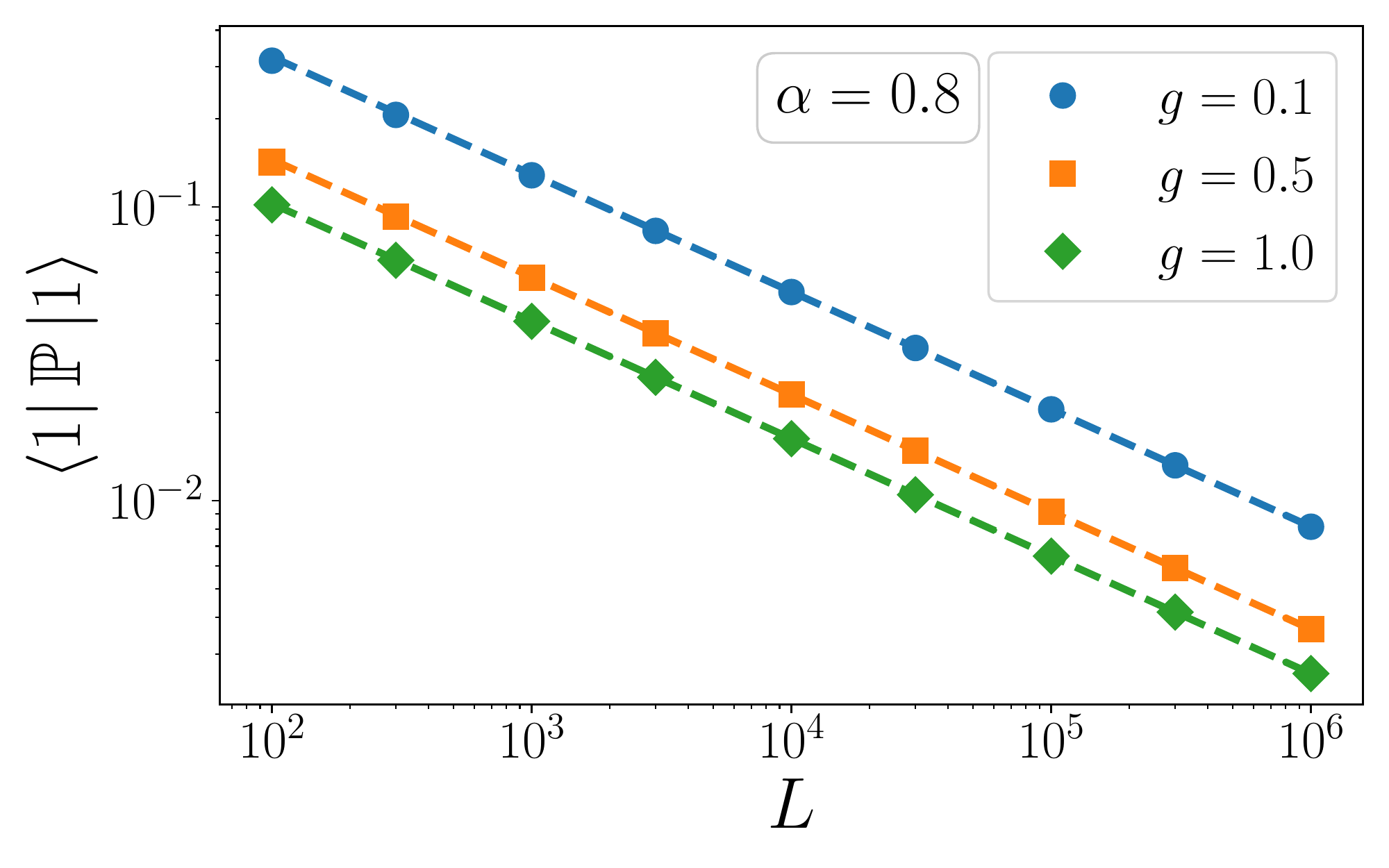}
 \includegraphics[width = .495\textwidth]{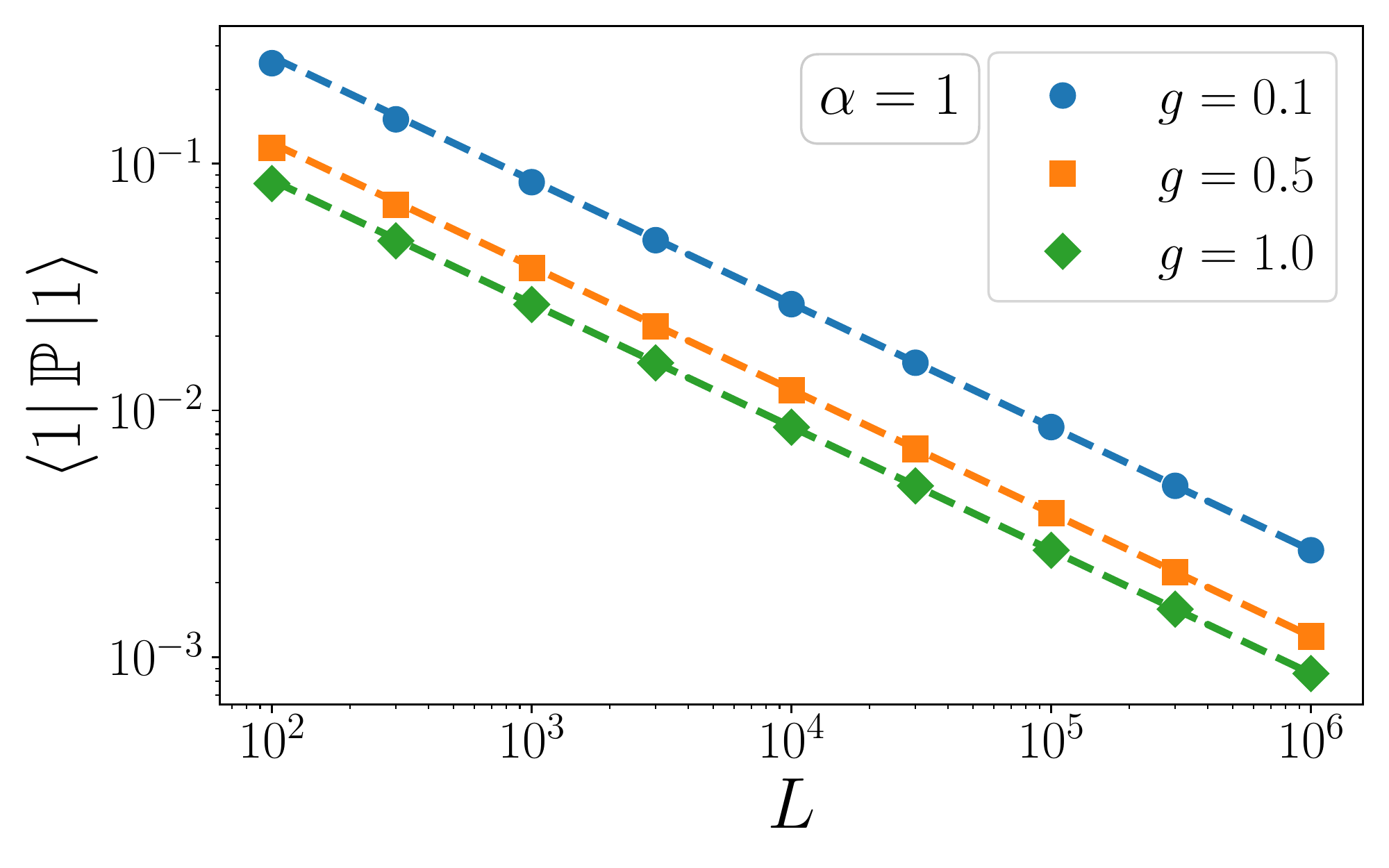}
 \includegraphics[width = .495\textwidth]{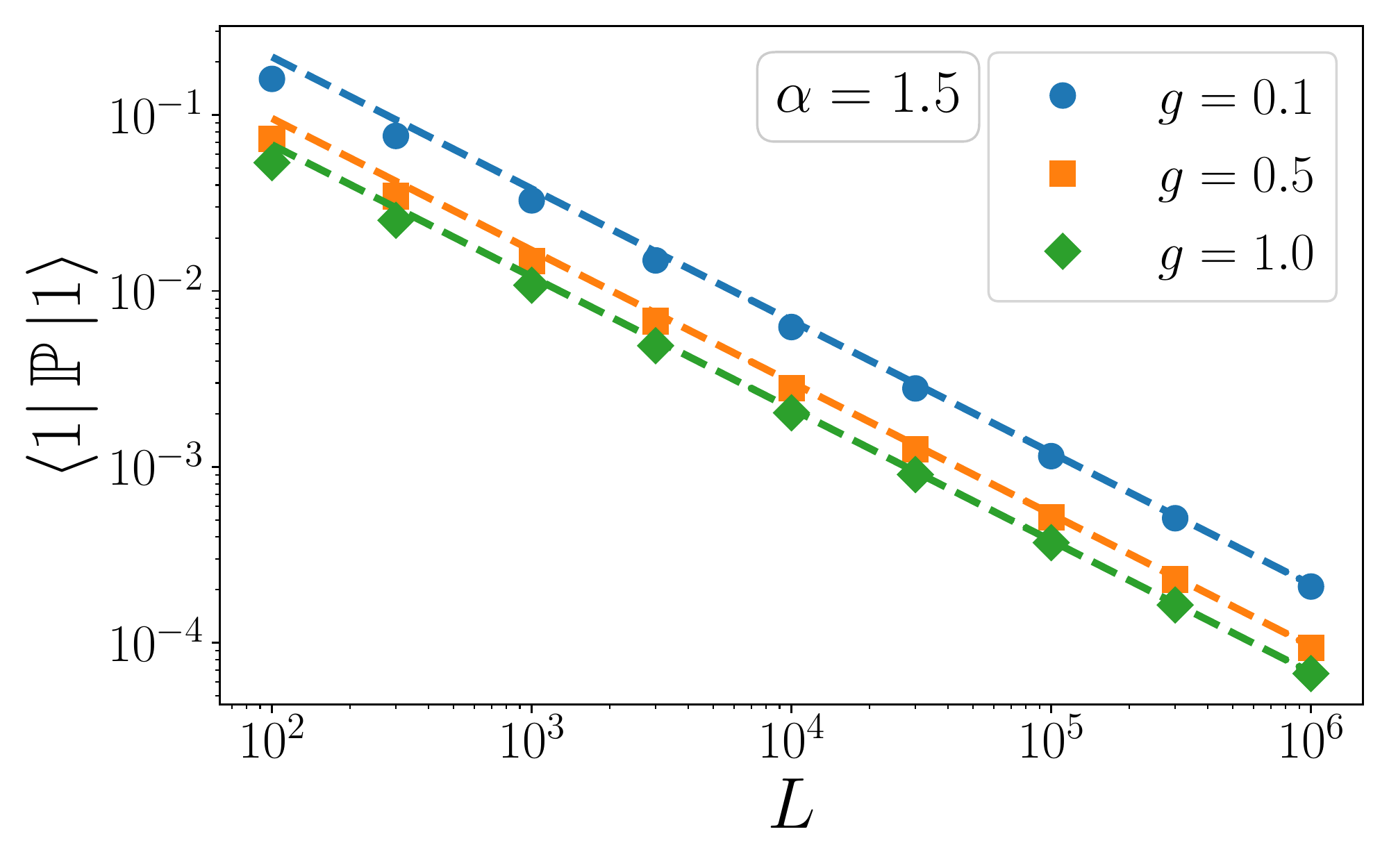}
 \caption{ Finite-size scaling behavior of $\chi_A^t=\langle1|\mathbb{P}|1\rangle_A$ in the 
 ordered phase of the QSM with long-range interactions. We plot $\chi_A^t$ versus $L$. Notice the logarithmic scale on both axes. The different panels correspond to 
 different values of the exponent $\alpha$ of the long-range interactions. In each figure 
 different symbols correspond to different values of $g$. The lines are the theory predictions 
 obtained summing~\eqref{eq:pth-fin} and~\eqref{eq:t3-fin}. 
}
\label{fig:flatP}
\end{figure}
%
%

\subsection{Finite-size contribution}
\label{app:p-fs}

Let us now determine the scaling behavior of the finite-size
contribution $\langle1|\mathbb{P}^{\mathrm{(L)}}|1\rangle_A$ (cf.~\eqref{eq:Pfs-def}). 
Specifically, here we have to evaluate a term $T_3$ of the form 
\begin{equation}
	\label{eq:t3-def}
	T_3 = \frac{2}{L} \sum_{j=1}^\infty \sum_{n,m=0}^{L/2} 
	e^{\II (n-m)k} \cos(L j k) \sqrt{2\mu + \omega(k)}.
\end{equation}
First, we express $\cos(L j k)$  in terms of complex exponentials, and use the representation
\begin{equation}
	e^{\II k x} = \int_\gamma \frac{\D s}{2\pi \II} (-\II x)^{-s} \Gamma(s), 
\end{equation}
which is  the analog of~\eqref{eq:cos}. 
Again, the path $\gamma$ is chosen as in~\eqref{eq:cos}, and it encloses 
the whole negative real axis. Subsequently, we 
exploit that the double sum in~\eqref{eq:t3-def} only depends on $n-m$. 
We can use~\eqref{eq:double-sum} and~\eqref{eq:u-int} to obtain 
\begin{equation}
	\label{eq:t3-res}
	T_3 = \frac{1}{2} \sum_{j=1}^\infty 
	\sum_{q=-L/2}^{L/2} \int_\gamma \frac{\D s}{2\pi \II}
	\Gamma(s) \frac{(-\II)^s}{(q\pm L j)^s} \left(1-\frac{2}{L}|q|\right) \mathfrak{J}(s), 
\end{equation}
where one has to  sum over the $\pm$. 
We can neglect the term with $q=0$ in~\eqref{eq:t3-res} because it is 
subleading. We can also  combine the contributions for $q$ and 
$-q$ in the sum. After using the same contour  $\gamma'$ as in~\eqref{eq:t1-3}, 
and after performing the sum over $j$, 
we obtain 
\begin{equation}
 T_3 \simeq \sum_{q=1}^{L/2} \int_\gamma \frac{\D s}{2\pi \II} 
 \Gamma(s) \cos\left(\frac{\pi}{2} s\right) L^{-s}\zeta\left( s,1\pm\frac{q}{L} \right)\left(1-\frac{2}{L}|q|\right) \mathfrak{J}(s).
\end{equation}
Again, the leading scaling behavior in the limit $L\to\infty$ is given by 
the residue at $s=1+\alpha/2$. We obtain 
\begin{equation}
	\label{eq:t3-fin}
	\langle1|\mathbb{P}^{\rm(L)}|1\rangle_A \simeq \Gamma\left(1+\frac{\alpha}{2}\right)\cos\left(1+\frac{\alpha}{2}\right)
 \int_0^1 (1-x) \zeta\left( 1+\frac{\alpha}{2},1\pm\frac{x}{2} \right)\frac{\D x}{2\pi} L^{-\frac{\alpha}{2}}
\end{equation}
where one has to sum over the signs in the argument of the Hurwitz zeta function, and 
we replaced the sum over $q$ with an integral. Importantly, the finite size contribution 
to $\chi_A^t$ is ${\mathcal O}(L^{-\alpha/2})$, as the thermodynamic one 
(cf.~\eqref{eq:pth-fin}).

\section*{References}
\bibliographystyle{iopart-num.bst}
\bibliography{bibliography}

\end{document}